\documentclass[acmsmall,screen,nonacm]{acmart}

\AtBeginDocument{%
  }

\setcopyright{acmlicensed}
\copyrightyear{2025}
\acmYear{2025}
\acmArticleType{Review}
\acmCodeLink{https://github.com/NoorKhalal/Scarse-Labels-in-CVD}
\acmJournal{CSUR}

\usepackage{placeins}
\usepackage[most]{tcolorbox}

\usepackage{listings}
\usepackage{subcaption}
\usepackage{hyperref}

\usepackage{tabularx}
\usepackage{array}

\usepackage{tcolorbox}
\definecolor{acmblue}{HTML}{0076A8}

\newtcolorbox{rqsummary}[1]{
    colback=acmblue!10,
    colframe=acmblue,
    colbacktitle=acmblue,
    coltitle=white,
    title={#1},
    fonttitle=\bfseries,
    boxrule=0.7pt,
    left=6pt, right=6pt, top=6pt, bottom=6pt,
}

\usepackage{booktabs}
\usepackage{tabularx}
\usepackage{pifont}      
\usepackage{makecell}    
\usepackage{array}       
\usepackage{wasysym} 

\newcommand{\cmark}{\ding{51}} 
\newcommand{\xmark}{\ding{55}} 
\newcommand{\pmark}{\LEFTcircle} 
\newcommand{\low}{L}
\newcommand{\med}{M}
\newcommand{\high}{H}

\begin{document}


\title{When Labels Are Scarce: A Systematic Mapping of Label‑Efficient Code Vulnerability Detection}

\author{Noor Khalal}
\orcid{0009-0008-9885-8166}
\email{noor.khalal@etu.u-paris.fr}
\affiliation{%
  \institution{Paris Cité university}  
  \country{France}
}
\affiliation{%
  \institution{CDC Informatique}
  \city{Paris}
  \country{France}
}

\author{Chakib Fettal}
\orcid{0000-0001-7684-5569}
\affiliation{%
  \institution{CDC Informatique}
  \city{Paris}
  \country{France}}
\email{chakib.fettal@caissedesdepots.fr}

\author{Lazhar Labiod}
\orcid{0000-0001-8641-8050}
\email{lazhar.labiod@u-paris.fr}
\affiliation{%
  \institution{Paris Cité university}
  \city{Paris}
  \country{France}
}

\author{Mohamed Nadif}
\orcid{0000-0002-0007-3950}
\email{mohamed.nadif@u-paris.fr}
\affiliation{%
  \institution{Paris Cité university}
  \city{Paris}
  \country{France}
}

\renewcommand{\shortauthors}{Khalal et al.}

\begin{abstract}
Machine-learning–based code vulnerability detection (CVD) has progressed rapidly, from deep program representations to pretrained code models and LLM-centered pipelines. Yet dependable vulnerability labeling remains expensive, noisy, and uneven across projects, languages, and CWE types, motivating approaches that reduce reliance on human labeling. This survey maps these approaches, synthesizing five paradigm families and the mechanisms they use. It connects mechanisms to token, graph, hybrid, and knowledge-based representations, and consolidates evaluation and reporting axes that limit comparison (label-budget specification, compute/cost assumptions, leakage, and granularity mismatches). A Design Map and constraint-first Decision Guide distill trade-offs and failure modes for practical method selection.
\end{abstract}

\ccsdesc[500]{Security and privacy~Intrusion/anomaly detection and malware mitigation}
\ccsdesc[500]{Computing methodologies~Learning paradigms}

\keywords{vulnerability detection, semi-supervised learning, self-supervised learning,
positive–unlabeled learning, contrastive learning, few-shot learning, domain adaptation, LLMs, code graphs}


\settopmatter{printacmref=false}
\maketitle

\section{Introduction}

Code Vulnerability Detection (CVD) lies at the intersection of software engineering and
cybersecurity. Its goal is to identify security-relevant defects early, preferably before they become
exploitable in deployed systems. Traditional vulnerability detection approaches rely on static analysis (e.g., taint
tracking, data-flow and control-flow reasoning, pattern matching) or dynamic analysis (e.g.,
coverage-guided fuzzing and symbolic execution). These methods remain indispensable components of
modern security pipelines, yet they face persistent challenges: high false-positive rates,
project-specific heuristics that do not generalize well across languages or ecosystems, and
nontrivial computational overhead during large-scale analysis. These limitations are widely noted in
recent surveys~\cite{research_107_2022_survey_data_driven_cvd,survey_007_2025_survey_source_code_vul_analysis_based_dl,survey_008_2024_comprehensive_review_cubersecurity_vuldetect_methodologies}.

Over the past decade, a growing body of work has reframed CVD as a learning problem. Early
deep-learning approaches such as VulDeePecker~\cite{research_082_2018_Vuldeepecker} and
Devign~\cite{Dataset_002_2019_Devign} introduced neural models over code gadgets, sequences, and
program-graph structures, inspiring a long line of GNN-based methods summarized in
recent surveys~\cite{research_178_Survey_VulPred_using_GNNs, research_179_Survey_sourceCVD_GNNs,
survey_007_2025_survey_source_code_vul_analysis_based_dl, research_066_2025_A_systematic_mapping_study_on_graph_machine_learning_for_static_source_code_analysis,
research_115_survey_cvd_ml_data_mining, research_132_taxonimy_software_vulnerability_ml_approaches}.
Concurrently, pretrained code models such as CodeBERT~\cite{research_181_CodeBert},
GraphCodeBERT~\cite{research_182_GraphCodeBert}, and CodeT5~\cite{research_180_CodeT5} demonstrated
that large-scale representation learning can significantly improve function-level detection and Common Weakness Enumeration (CWE)-type classification~\cite{research_183_CodeBert_framework_evaluating_classification_vulpred,
research_089_2025_fusing_language_models_and_online_distilled_graph_neural_networks_for_cvd_VulLMGGNNs}.

Most recently, LLMs have accelerated interest in prompting, in-context examples, and
parameter-efficient fine-tuning for CVD. Controlled evaluations report competitive performance under carefully
engineered prompts~\cite{research_110_llm4cvd_short_results_directions}, while other studies highlight instability,
false positives, and sensitivity to context size, formatting, or task design~\cite{research_108_2023_cvd_llms}.
Follow-up analyses~\cite{research_185_llm_vuldetec_emerging_results_2024} show that LLMs can benefit from
context structuring, retrieval, or human-in-the-loop pipelines. Recent surveys further emphasize persistent
challenges, including domain shift, cross-language generalization, robustness, and explainability
\cite{research_030_2024_Prompt_Enhanced_Software_Vulnerability_Detection_Using_ChatGPT,
research_032_2024_LLMs_cannot_Reliably_Identify_and_Reason_about_Security_Vulnerabilities_yet,
research_103_improving_cvd_llm_via_in_context_learning_information_fusion, survey_002_2025_LLMS_in_Software_Security_Vulnerability_Detection_Techniques_and_Insights}.

Despite this momentum, the supervision bottleneck persists. High-quality vulnerability labels
are expensive to obtain, unevenly distributed across CWE types, and prone to noise. Synthetic
benchmarks such as SARD/Juliet~\cite{Dataset_003_2017_Sard} provide valuable training corpora but do
not fully reflect real-world complexity. Curated vulnerability datasets (e.g., Big-Vul
\cite{Dataset_003_2020_BigVul}, Devign~\cite{Dataset_002_2019_Devign}, and
DiverseVul~\cite{Dataset_004_2023_Diversevul}) reveal pronounced distribution shifts across projects
and frequent label inconsistencies~\cite{research_186_limits_of_ml_in_cvd,
research_187_comprehensive_analysis_svd_dtasets}. The result is a fragmented landscape where
supervised methods remain dominant, but annotation cost, generalization, and compute constraints
motivate the exploration of \emph{label-efficient} alternatives.

\smallskip
\noindent\textbf{Positioning and scope.}
Existing CVD surveys provide valuable overviews of deep learning, graph-based modeling, and LLM-centric pipelines, but
they typically organize the literature by model family, representation type, or security capability, and treat label scarcity
as a recurring challenge rather than as a unifying synthesis axis. As summarized in
Table~\ref{tab:survey_positioning}, this leaves open a practical question for both researchers and practitioners:
\emph{given scarce labels, what families of methods are available, what mechanisms make them label-efficient, and under which
representation, dataset, granularity, and cost conditions are they comparable?}
To answer this question, we focus on empirical evidence reported in the CVD literature and adopt a
structured synthesis (rather than meta-analysis), because reporting of label budgets, splits, and compute is often incomplete
and heterogeneous (see Section~\ref{sec:threats_to_validity}). We operationalize this synthesis into actionable
guidance via the constraint-first \emph{Practitioner Decision Guide} (Table~\ref{tab:decision_guide}).

Beyond code-specific approaches, a wide range of paradigms aim to reduce reliance on labeled data.
In this survey, we organize the CVD literature around five recurring \emph{label-efficient families}:%
Semi-/Self-Supervised Learning, Contrastive Learning, Few-Shot/Meta-Learning, Prompt-Tuning, and LLM-based prompting and parameter-efficient adaptation;%
while tracking cross-cutting strategies such as domain adaptation, anomaly detection, retrieval, and human-in-the-loop
verification that can compose with multiple families. However, evidence remains scattered: papers often evaluate different
methods under incomparable conditions, mix heterogeneous representations (tokens, graphs, telemetry), assume different
label budgets, and report inconsistent compute and operational metrics. This fragmentation hinders cumulative progress and
creates a mismatch between reported ``state-of-the-art'' gains and deployable, budget-aware CVD practice.

\medskip
\noindent\textbf{Objective and Research Questions.}
Our goal is to synthesize label-efficient CVD through a reproducible, representation-aware lens that makes supervision,
data, and cost assumptions explicit. Concretely, we ask:

\begin{itemize}
    \item \textbf{RQ1:} Which label-efficient paradigm families have been applied to CVD, and what core mechanisms do they employ?
    \item \textbf{RQ2:} How do representation choices (tokens, graphs, hybrid designs) and related design decisions
    (task type, evaluation metrics, granularity and dataset characteristics) interact with these paradigm families?
    \item \textbf{RQ3:} How do studies specify and exploit label budgets, and what evidence exists regarding label efficiency?
    \item \textbf{RQ4:} What computational and operational costs do the proposed approaches entail?
    \item \textbf{RQ5:} What limitations, failure modes, and open challenges emerge across families, and what reporting omissions
    prevent verification and comparison?
\end{itemize}

\medskip
\noindent\textbf{Contributions.}
This survey makes four contributions.

(1) We provide, to our knowledge, the first structured mapping of label-efficient CVD methods organized by paradigm families
and core mechanisms (RQ1), and we distill their assumptions about labels, negatives, and supervision signals. Table~\ref{tab:survey_positioning}
positions our scope and synthesis axes relative to prior CVD surveys.

(2) We synthesize how representation choices and evaluation design shape feasibility at scale, linking token-, graph-, hybrid-,
and knowledge-based views to dataset families and code granularity choices (RQ2).

(3) We consolidate evidence and reporting patterns on label budgets, generalization, and compute/operational cost, and distill
recurring trade-offs and failure modes. We operationalize these findings into actionable design guidance via the \emph{Label-Efficient CVD Design Map}
(Table~\ref{tab:design_map}) and the deployment-constraint-oriented \emph{Practitioner Decision Guide} (Table~\ref{tab:decision_guide}).

(4) To support transparency, reproducibility, and future extensions, we release a
\href{https://github.com/NoorKhalal/Scarse-Labels-in-CVD/blob/main/}{\textcolor{acmblue}{publicly accessible repository}}
with the curated paper list, extracted metadata, and scripts used to generate all figures and tables in this study.

\medskip
\noindent The remainder of this paper is structured as follows. Section~\ref{sec:background} introduces label-efficient
learning concepts and the supervision spectrum in the context of CVD. Section~\ref{sec:study-design} outlines our study
design. Sections~\ref{sec:rq1}--\ref{sec:rq5} synthesize findings for each Research Question.
Section~\ref{sec:threats_to_validity} discusses threats to validity, and Section~\ref{sec:discussion}
derives a research agenda for future label-efficient CVD. Finally, Section~\ref{sec:conclusion} concludes the paper.

\section{Background}
\label{sec:background}

This section introduces the concepts needed to interpret label-efficient learning for CVD. We first outline the CVD task and the recurring challenges that
make supervised learning brittle or costly in practice. We then introduce the main label-efficient
paradigm families used in the CVD literature (later formalized in Section~\ref{sec:rq1} and summarized in
the Design Map, Table~\ref{tab:design_map}). Finally, we review the representation families that
CVD systems rely on and how they interact with label-efficient designs.

\subsection{Machine Learning for Code Vulnerability Detection}
\label{sec:bg-ml-cvd}

Vulnerability detection is a key step in the broader vulnerability management workflow, preceding
vulnerability analysis and remediation. Traditional detection commonly relies on (i) manual code
auditing, (ii) static analysis of source or binaries without execution, (iii) dynamic analysis during
execution, and (iv) fuzz testing to probe input-sensitive behaviors
\cite{DBLP:conf/uss/KimKRFTW0DX19, DBLP:conf/pldi/NethercoteS07}.
While these approaches remain essential, they can be labor-intensive, tool-dependent, and difficult
to scale across large and rapidly changing codebases.

Machine learning (ML) methods aim to complement these techniques by learning vulnerability signals
from data. Depending on the formulation, CVD tasks include binary classification (vulnerable vs.\
non-vulnerable), CWE-typed classification, and vulnerability localization at the function, block, or
line level. ML pipelines can ingest diverse evidence sources; including source code; derived program
structures, natural-language artifacts, and numerical/process telemetry; but are constrained by
label scarcity, label noise, distribution shift, and the operational cost of extracting structural
views
\cite{research_107_2022_survey_data_driven_cvd, research_132_taxonimy_software_vulnerability_ml_approaches}.

Granularity ranges from coarse-grained function/file-level prediction to fine-grained localization
and just-in-time (JIT) prediction at the commit or patch level. Granularity influences both the
available supervision and the representation requirements: localization tasks more often require
explicit modeling of control/data flow, whereas coarse-grained settings can sometimes be addressed
with token- or metric-based features
\cite{research_176_effort_aware_jit_defect_pred_unsup_could_be_better_than_sup}.

\paragraph{Core challenges motivating label efficiency.}
Five recurring challenges motivate label-efficient learning in CVD:
(i) \emph{label scarcity and noise}, as vulnerability annotation is costly and error-prone;
(ii) \emph{domain shift}, where models trained on one project or library generalize poorly to others
\cite{research_101_CVD_deep_domain_adaptation_max_margin_principle_cross_project};
(iii) \emph{program structure}, since many CWEs depend on long-range control or data-flow dependencies
\cite{research_164_multi_view_pretrained_model_cvd,
research_089_2025_fusing_language_models_and_online_distilled_graph_neural_networks_for_cvd_VulLMGGNNs};
(iv) \emph{compute and practicality}, as graph extraction and full model fine-tuning can be expensive
\cite{research_086_2024_SCL_CVD};
and (v) \emph{evaluation pitfalls}, including duplication-prone random splits and severe class
imbalance
\cite{research_176_effort_aware_jit_defect_pred_unsup_could_be_better_than_sup}.
Together, these constraints explain why purely supervised pipelines are often difficult to deploy
and why CVD is a natural testbed for label-efficient designs.

\subsection{Label-Efficient Learning Paradigm Families in CVD}
\label{sec:background-paradigms}

Across the CVD literature, label efficiency is achieved through a small set of recurring
\emph{paradigm families}. These families are best viewed as an organizing ontology rather than a
strict taxonomy: boundaries can be fuzzy, and many systems combine multiple mechanisms within a
single pipeline. We use these families to structure the detailed synthesis in Section~\ref{sec:rq1}
and summarize typical design patterns and trade-offs in the Design Map (Table~\ref{tab:design_map}).

\paragraph{Semi-supervised learning.}
Semi-supervised methods exploit unlabeled code alongside a small labeled set, commonly via
pseudo-labeling, teacher--student training, or consistency regularization.

\paragraph{Self-supervised representation learning.}
Self-supervised methods pretrain on unlabeled code using proxy objectives (e.g., masked prediction),
producing representations that improve sample efficiency for downstream vulnerability prediction
\cite{research_181_CodeBert, research_220_2019_Pretrain_graph_neural_networks}.

\paragraph{Contrastive learning.}
Contrastive objectives learn invariances by aligning related views and separating non-matching
examples. In CVD, contrastive learning appears both in pretraining and in label-efficient fine-tuning
(e.g., supervised-contrastive variants), and is often sensitive to view/augmentation and sampling
design.

\paragraph{Few-shot and meta-learning.}
Few-shot and meta-learning aim to adapt to new CWE types or domains with only a handful of labeled
examples, typically through episodic training and metric- or optimization-based adaptation
\cite{research_159_survey_fsl, research_224_2021_Meta_Learning_NN_survey}.

\paragraph{Prompting and parameter-efficient adaptation.}
LLM-based approaches reduce supervision needs via prompting and lightweight adaptation (e.g., LoRA,
adapters), shifting effort from collecting labels to prompt/context construction and efficient tuning.

\subsection{Code Representations for Vulnerability Detection}
\label{sec:bg-representations}

The effectiveness and practicality of label-efficient paradigms depend strongly on the code
representation. In our corpus (RQ2), four recurring representation families appear in CVD systems.

\paragraph{Token-based representations.}
Token-based approaches treat code as a sequence of lexical units or sub-tokens, making them
compatible with pretrained code language models and LLM prompting pipelines. They naturally support
self-supervised pretraining and parameter-efficient adaptation under limited supervision.

\paragraph{Graph and flow representations.}
Graph-based encodings explicitly model program structure. Abstract syntax trees (ASTs) capture
syntax; control-flow graphs (CFGs) encode execution paths; program dependence graphs (PDGs) model
control and data dependencies; and code property graphs (CPGs) integrate multiple views
\cite{research_68_2014_CPG_Code_Property_Graphs_joern}.
These representations are well suited for vulnerabilities that depend on long-range flows and are
often paired with structure-aware objectives (including contrastive and semi-supervised designs).

\paragraph{Hybrid (multi-view) representations.}
Hybrid approaches fuse token embeddings with structural views (e.g., token+graph), or combine
multiple learned views to improve robustness across projects and CWE categories
\cite{research_089_2025_fusing_language_models_and_online_distilled_graph_neural_networks_for_cvd_VulLMGGNNs}.

\paragraph{Knowledge-based representations.}
Knowledge-based representations augment learned signals with externally defined features such as
static-analysis warnings, handcrafted metrics, vulnerability patterns, or commit/process metadata.
They are frequently used in conjunction with token or hybrid models to inject domain priors when
labels are limited.

The remainder of this survey (Sections~\ref{sec:rq1}--\ref{sec:discussion}) examines
how these paradigm and representation choices interact in practice, and Table~\ref{tab:design_map}
summarizes the resulting design patterns, cost considerations, and common failure modes.

\section{Study Design and Protocol}
\label{sec:study-design}

\subsection{Motivation and Scope}

Prior surveys on CVD and related software-security settings primarily organize the literature by model family,
representation, or, more recently, LLM capabilities. Table~\ref{tab:survey_positioning} situates this work relative to closely related surveys and overviews.
Existing reviews largely organize the field by model/representation families or provide LLM-centric summaries, and typically treat annotation scarcity as a challenge rather than a primary synthesis axis.
In contrast, we structure the mapping around supervision regimes and label-budget reporting, enabling a supervision-focused analysis of mechanisms and evidence across CVD studies.

\begin{figure}[h]
    \centering
    \includegraphics[width=\linewidth]{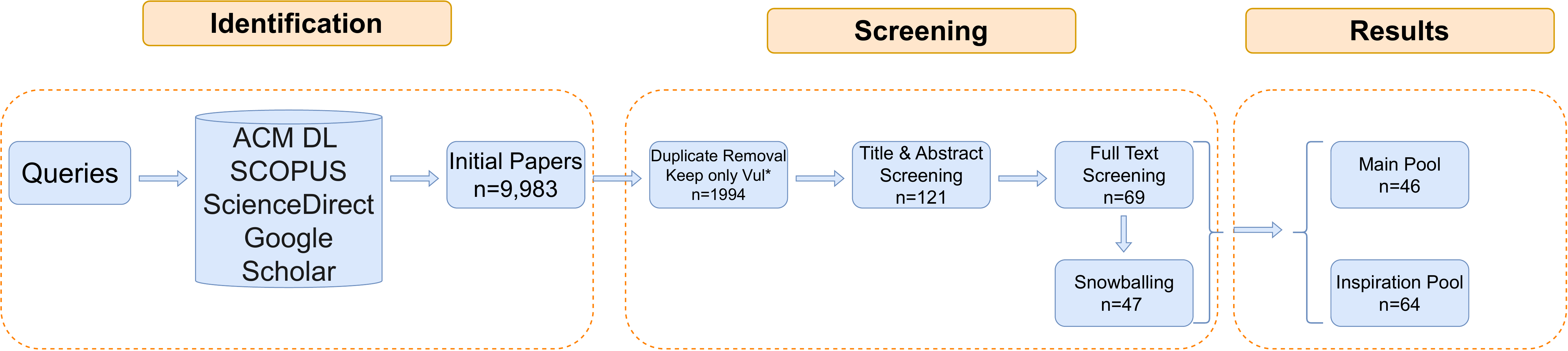}
    \caption{Study selection flow.}
    \label{fig:prisma-flow}
\end{figure}

We followed established guidelines for systematic literature reviews and mapping studies in software engineering to design the
search, screening, and extraction protocol~\cite{kitchenham2007guidelines,petersen2015guidelines,miscellaneous_000_2007_Guidelines_for_performing_systematic_literature_reviews_in_software_engineering}.
We report study identification and selection in a flow diagram for transparency (Figure~\ref{fig:prisma-flow}), and release
extracted study-level metadata and coding artifacts to support inspection and reuse\footnote{\url{https://github.com/NoorKhalal/Scarce-Labels-in-CVD}}.

\begin{tcolorbox}[colback=white,colframe=black,boxrule=0.5pt,arc=1pt,left=4pt,right=4pt,top=3pt,bottom=3pt]
\small\textbf{Search keywords.}
\begin{itemize}
  \setlength{\itemsep}{1pt}
  \setlength{\topsep}{2pt}
  \item \textit{Core CVD terms:} \texttt{vulnerability detection}, \texttt{software vulnerability}, \texttt{source code security}, \texttt{static code analysis}.
  \item \textit{Label-efficiency terms:} \texttt{self-supervised}, \texttt{semi-supervised}, \texttt{contrastive}, \texttt{positive-unlabeled/PU}, \texttt{few-shot}, \texttt{meta-learning}, \texttt{active learning}, \texttt{domain adaptation/transfer}, \texttt{prompt*}, \texttt{in-context}, \texttt{PEFT}.
  \item \textit{Representation terms:} \texttt{AST}, \texttt{CFG}, \texttt{PDG}, \texttt{CPG}, \texttt{graph}.
\end{itemize}
\end{tcolorbox}

\begin{table}[h]
\centering
\scriptsize
\setlength{\tabcolsep}{3.5pt}
\renewcommand{\arraystretch}{1.15}
\caption{Positioning relative to closely related surveys and overviews. ``limited'' indicates the topic is discussed but not used as a primary organizing axis; ``\pmark'' indicates partial coverage ; ``--'' indicates the attribute is not clearly stated in the available abstract/metadata used for coding.}
\label{tab:survey_positioning}
\begin{tabularx}{\linewidth}{l c X c c c c}
\toprule
\textbf{Work} & \textbf{Year} & \textbf{Primary organization / scope} & \textbf{CVD} & \textbf{LLMs} & \textbf{Label-eff.\ axis} & \textbf{Budget lens} \\
\midrule

\cite{survey_006_2024_sys_lit_review_automated_vuldetec_using_ml} & 2025 &
Systematic literature review of ML/DL for automated software vulnerability detection; organizes via multiple RQs (models, datasets, representations, tools, CWE coverage). &
\cmark & \pmark & \xmark & limited \\

\cite{survey_007_2025_survey_source_code_vul_analysis_based_dl} & 2025 &
Survey of deep learning for (static) source-code vulnerability detection; taxonomy centered on source-code representations (token/graph) and DL variants. &
\cmark & \pmark & \xmark & \xmark \\

\cite{survey_002_2025_LLMS_in_Software_Security_Vulnerability_Detection_Techniques_and_Insights} & 2025 &
Survey of LLM-based vulnerability detection (problem formulation, model selection, datasets, evaluation, challenges). &
\cmark & \cmark & \xmark & limited \\

\cite{survey_005_2025_VulDetection_by_LLM_FormalVerification_Hybrid} & 2025 &
Overview from formal verification to LLMs and hybrid approaches (broad methods landscape; not supervision-regime centric). &
\cmark & \cmark & \xmark & \xmark \\

\cite{survey_008_2024_comprehensive_review_cubersecurity_vuldetect_methodologies} & 2024 &
Broad cybersecurity vulnerability detection methodologies (beyond ML/DL; multiple detection paradigms). &
\pmark & \pmark & \xmark & \xmark \\

\cite{survey_003_2024_LLMS4SE_Systematic_Literature_Review} & 2024 &
Broad SLR of LLMs for software engineering tasks (models, data, optimization/evaluation, and SE applications). &
\xmark & \cmark & \xmark & \xmark \\

\cite{survey_013_2024_Survey_Semi_supervised_cybersecurity} & 2024 &
Survey of semi-supervised learning in cyber-security (label scarcity as motivation; emphasis on intrusion detection and related domains). &
\xmark & \xmark & \cmark & limited \\

\cite{research_110_llm4cvd_short_results_directions} & 2024 &
Short overview (NIER) of emerging results and directions for LLM-based vulnerability detection. &
\cmark & \cmark & \xmark & \xmark \\

\cite{research_132_taxonimy_software_vulnerability_ml_approaches} & 2021 &
Taxonomy of vulnerability detection and ML approaches; emphasizes trends across methods/features/datasets and includes a high-level discussion of semi-/transfer learning. &
\cmark & \xmark & \xmark & Limited \\

\cite{survey_014_2020_Literature_survey_of_deep_learning_based_vulnerability_analysis_on_source_code} & 2020 &
Early survey of deep learning-based vulnerability analysis on source code; systematizes by objectives/focus/features/architectures and trends. &
\cmark & \xmark & \xmark & \xmark \\

\cite{research_115_survey_cvd_ml_data_mining} & 2017 &
Survey of software vulnerability analysis/discovery using machine-learning and data-mining; discusses categories, challenges, and open areas (broad, pre-LLM). &
\cmark & \xmark & \xmark & \xmark \\

\midrule
\textbf{This work} & 2026 &
Systematic mapping of \emph{label-efficient} paradigms for CVD: supervision regimes and mechanisms (semi-/self-supervised, PU/weak supervision, few-shot/meta-learning, contrastive, LLM prompting/PEFT), with an explicit label-budget reporting lens (RQ3/RQ5). &
\cmark & \cmark & \cmark & \cmark \\
\bottomrule
\end{tabularx}
\end{table}

\subsection{Information Sources and Time Window}
We queried Google Scholar, ACM Digital Library, Scopus, and ScienceDirect, which collectively index the venues most
active in CVD research (e.g., ICSE, FSE). Searches were last executed in October~2025.
We considered English-language peer-reviewed and preprint studies with sufficient methodological detail for extraction.
The coverage window spans January~2006 to September~2025, capturing early semi-/unsupervised work in software engineering
through LLM-era label-efficient methods.

\subsection{Search Strategy and Study Selection}
We used an iterative search strategy, starting from broad CVD queries and progressively adding label-efficiency terms as
early screening and snowballing revealed additional relevant mechanisms and terminology. To maintain high precision for
the target task, we applied a conservative first-pass title heuristic requiring \texttt{vulnerab*}, followed by
abstract/introduction screening and full-text assessment.

\paragraph{Eligibility criteria.}
We included studies that satisfied all of the following:
\begin{itemize}
  \item[\textbf{IC1}] The primary task is CVD, or a directly transferable security-relevant code-analysis task.
  \item[\textbf{IC2}] The approach uses a non-fully supervised paradigm as a core methodological element, or reports explicit baselines enabling comparison.
  \item[\textbf{IC3}] The study provides sufficient methodological detail to extract the supervision regime, representation, data/split protocol, and evaluation setup.
  \item[\textbf{IC4}] The paper is accessible in English (peer-reviewed or preprint).
\end{itemize}

We excluded studies matching any of the following:
\begin{itemize}
  \item[\textbf{EC1}] Surveys, tutorials, position papers, or non-research reports (no extractable empirical protocol).
  \item[\textbf{EC2}] Non-SE domains or tasks not transferable to our supervision-focused CVD analysis.
  \item[\textbf{EC3}] Learning-free pipelines (e.g., static/dynamic analysis without ML), or dynamic exploitation/fuzzing-only setups without an ML component.
  \item[\textbf{EC4}] Inaccessible or unextractable reports, or insufficient methodological clarity for coding.
\end{itemize}

These criteria were applied in the title/abstract screening and full-text assessment stages reflected in Figure~\ref{fig:prisma-flow}.
Paradigm families used to organize the synthesis (RQ1) were derived from the included corpus during screening and coding, rather than predefined a priori.

\paragraph{Selection stages and corpus counts.}
As summarized in Figure~\ref{fig:prisma-flow}, database searching returned $9{,}983$ records. After de-duplication, we applied a precision-oriented \texttt{vulnerab*} filter at the title and abstract level to retain high-relevance candidates, resulting in $1{,}994$ records.
 We then performed title-and-abstract screening (full reading of both fields), which yielded $121$ candidates
for full-text review. Full-text assessment retained $69$ studies that met our eligibility criteria.
Backward/forward snowballing from these anchors added $47$ additional relevant studies. The final corpus comprises
$46$ Main-set CVD studies and $64$ Inspiration-set studies (defined below).

\paragraph{Main vs.\ Inspiration corpora.}
To answer RQ1--RQ5 while preserving task fidelity, we formed two disjoint corpora:
(i) a \textbf{Main Set} of studies whose primary task is CVD and that employ a non-fully supervised regime as a central contribution; and
(ii) an \textbf{Inspiration Set} of adjacent SE studies instantiating the same paradigms on related tasks (e.g., defect prediction).
Only the Main Set informs the core synthesis and any performance-related discussion; the Inspiration Set is used to contextualize mechanisms
and transferability without aggregating results across tasks.

\subsection{Data Extraction and Coding}
Screening and data extraction followed a protocol agreed by the author team.
Co-authors reviewed the eligibility criteria and extraction schema and cross-checked a subset of studies; discrepancies were resolved through discussion and protocol refinement.

For each included study, we extracted bibliographic metadata; task granularity; supervision regime; representation;
datasets and split strategies; label-budget reporting; evaluation metrics; compute/cost proxies; and artifact availability.
Coded fields were stored in a structured schema to support reproducible synthesis and automated figure generation.
Because supervision is reported heterogeneously (e.g., labeled fractions, $K$-shot settings, positives-only regimes), we use an operational definition of label efficiency (Section~\ref{sec:label_eff_definition}) to support consistent interpretation in RQ3 and RQ5.
The extraction schema was refined iteratively as new mechanisms and representations emerged; updates were applied consistently across previously coded entries.

\subsection{Operational Definition of Label Efficiency}
\label{sec:label_eff_definition}
We adopt an operational view: a method is label-efficient if it reaches a desired effectiveness level using fewer labeled instances under a fixed dataset, split protocol, and evaluation metric.
Throughout the survey, label budget may be expressed as a labeled fraction, a $K$-shot setting, positives-only supervision, or an annotation-effort proxy; these budget types are not directly interchangeable without additional protocol detail.
We use this definition as a \emph{comparability lens} to categorize how studies report budgets and to motivate a minimum reporting checklist (RQ5); we avoid curve-based meta-analysis when primary studies do not report standardized budget-performance curves.

\subsection{Synthesis Plan}
We conduct a structured narrative synthesis organized by paradigm family. For each paradigm, we characterize the learning signals and supervision mechanisms, relate them to representations, and summarize how studies report label budgets and compute/cost proxies.
When experimental setups are comparable, we align baselines and report relative outcomes; otherwise, we avoid quantitative aggregation that would imply meta-analysis.

\subsection{Replication Artifacts and Deviations}
We release extracted metadata and coding fields, along with scripts used to aggregate results and generate figures, in our repository:
\href{https://github.com/NoorKhalal/Scarce-Labels-in-CVD/tree/main}{\textcolor{acmblue}{Scarce-Labels-in-CVD}}.
Two protocol refinements occurred during the review: (i) iterative expansion of search keywords as additional paradigms emerged during early screening; and (ii) introduction of the Inspiration Set when snowballing identified high-quality instances of target paradigms on adjacent tasks.
Threats to validity and residual biases (e.g., split design, duplication/leakage controls, compute reporting gaps) are discussed in Section~\ref{sec:threats_to_validity} dedicated to the Threats to Validity.

\section{RQ1: Taxonomy of Label‑Efficient Paradigms}
\label{sec:rq1}

In our first RQ, we ask \emph{which label-efficient paradigms have been applied to CVD, and through which mechanisms do they reduce reliance on annotated vulnerability labels?}
We organize the answer around five recurring paradigm families in the CVD literature: semi-/self-supervised learning, contrastive learning, few-shot/meta-learning, prompt tuning, and LLM-based approaches.

We use \emph{family} pragmatically to denote a dominant learning regime (i.e., the primary source of supervision or adaptation signal), rather than a mutually exclusive label: many systems combine multiple regimes within a single pipeline.
We therefore treat this taxonomy as a primary lens to organize discussion, not as a claim that papers belong to exactly one box. Concretely, we describe each family by its core label-efficiency mechanism, then highlight common hybridizations that recur across papers.

Within each family, we (i) define the paradigm at the level needed to interpret CVD papers, (ii) summarize how it is instantiated in CVD, and (iii) highlight transferable design choices that recur across studies.
We additionally separate a set of \emph{cross-cutting strategies} (e.g., domain adaptation, anomaly detection, active learning) that can be composed with multiple families and are therefore best treated as orthogonal design choices rather than stand-alone families.

\paragraph{Role of Inspiration Studies.}
RQ1 is formulated at the level of \emph{label-saving mechanisms} (how supervision is reduced), rather than at the level of a specific benchmark or dataset. We therefore adopt a two-corpus evidence strategy: (i) the \textbf{Main pool} establishes which mechanisms have been instantiated
\emph{within CVD}; (ii) the \textbf{Inspiration pool} captures mechanisms that are mature in closely related software-engineering tasks under comparable supervision constraints (e.g., defect prediction, clone detection, anomaly detection) and are technically compatible with CVD pipelines.
This separation improves taxonomy completeness while preserving construct validity: Inspiration studies are used only for mechanism characterization and design transfer, and are not combined with the Main pool in performance summaries or CVD-specific claims.

\begin{figure}[h]
    \centering
    \includegraphics[width=\textwidth]{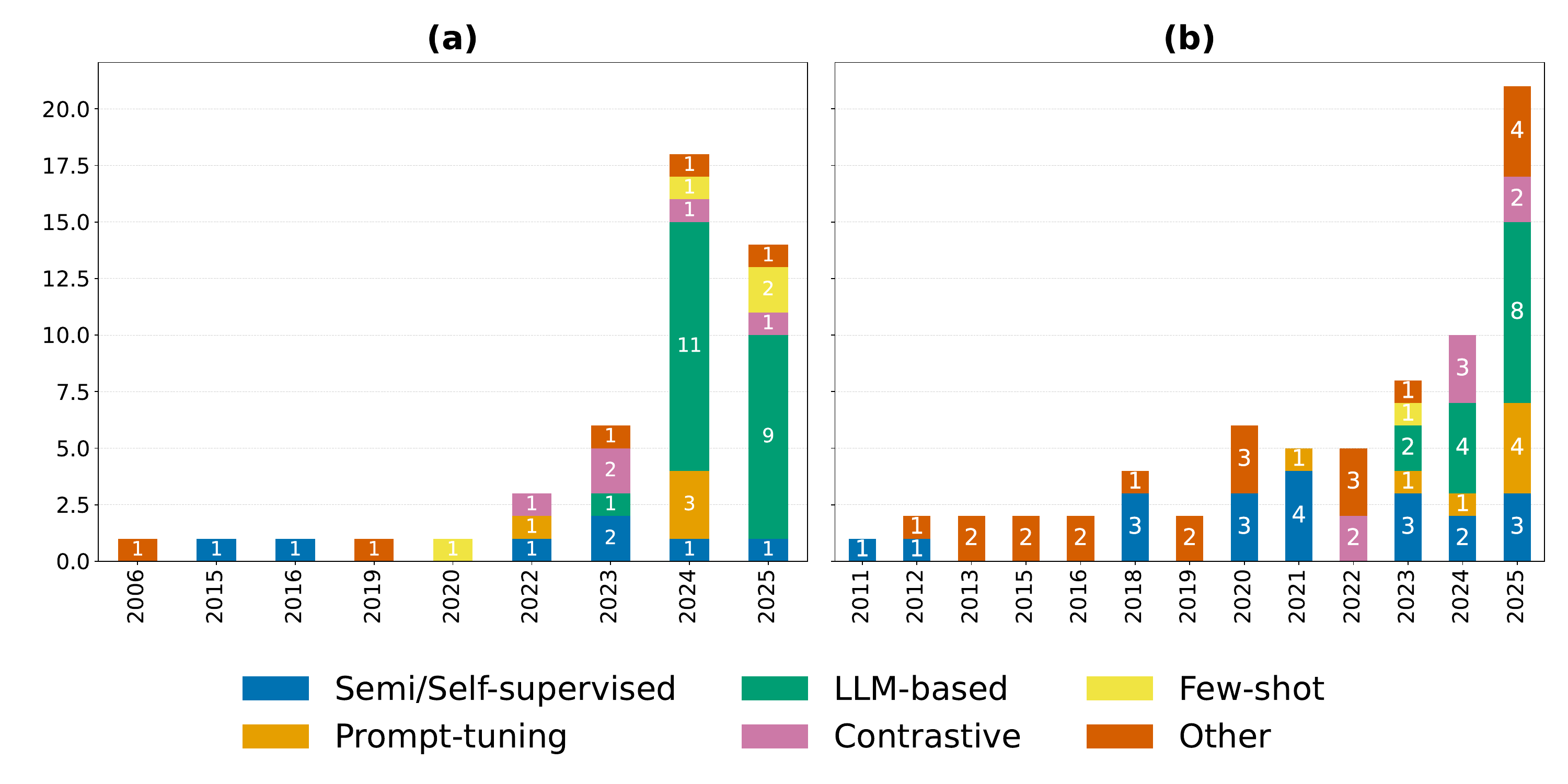}
    \caption{Annual publication counts of included studies by paradigm family for the Main pool (a) and the Inspiration pool (b). ``Other'' refers to Cross-cutting mechanisms and specialized settings.}
    \Description{Two stacked bar charts showing annual counts of studies by paradigm family for the Main pool (a) and the Inspiration pool (b). }
    \label{fig:year_distribution}
\end{figure}

Figure~\ref{fig:year_distribution} contextualizes the evidence base for RQ1. The Main pool contains isolated early instances but grows sharply in recent years, driven primarily by LLM-based and prompt-centric approaches.
The Inspiration pool is larger and spans a longer period, showing that several label-saving mechanisms were explored earlier in adjacent SE tasks before being adopted in CVD. We use this separation to distinguish (i) mechanisms demonstrated in CVD from (ii) mechanisms that are transferable but not yet
widely validated in CVD.

\subsection{RQ1-A: Semi- and Self-Supervised Learning}
\label{subsec41:semi-self}
Semi-supervised learning combines a small labeled seed with a large unlabeled pool through mechanisms such as pseudo-labeling, consistency regularization, and graph-based propagation. Self-supervised learning constructs synthetic targets (contrastive or masked reconstruction) to pretrain representations later adapted with few labels.

\textbf{Instantiation for CVD.}
Early semi-supervised work couples engineered analysis features with disagreement or graph-driven utilization of unlabeled data. Shar et al.~\cite{research_146_web_app_vul_pred_hybrid_program_analysis_ml} use interprocedural slices to supply information-flow, control-flow, and syntax-derived attributes; an ensemble of decision-tree learners is then co-trained to self-label unlabeled statements only when a committee agrees with high confidence, gradually densifying supervision without manual annotation. By contrast, Meng et al.~\cite{research_142_predicting_buffer_overflow_using_semi_supervised_learning} build a similarity graph over functions from AST-derived attributes (index/length patterns and loop/control indicators) and propagate labels harmonically from a handful of seeds; their mechanism emphasizes manifold smoothness rather than committee consensus.

Modern pipelines move supervision from feature engineering to representation learning. Yu et al.~\cite{research_74_2025_less_is_more_semi_supervised_deep_learning_vuldetec} instantiate a teacher–student loop atop strong base encoders (LineVul~\cite{research_027_2022_LineVul}; a token Transformer, and ReVeal~\cite{research_084_2021_DL_based_VulDete_are_we_there_yet_ReVeal}; a graph-based model), but filter pseudo-labels through certainty-aware sampling and train with noise-robust cross-entropy and triplet objectives; the effect is to amplify reliable signals from the unlabeled pool while discouraging confirmation errors in ambiguous regions of the space. Wen et al.~\cite{research_73_2023_less_is_enough_positive_unlabeled_model_for_vuldetec} pursue a complementary route using PU: starting from the assumption that only vulnerabilities (positives) are explicitly labeled, they select high-confidence unlabeled items via a distance-aware criterion and interleave progressive fine-tuning with mixed supervision losses so the model learns a calibrated decision boundary without ever observing “true” negatives.

Self-supervision supplies label-free representations that transfer to downstream detection with few labels. Cheng et al.~\cite{research_124_path_sensitive_embedding_contrastive_learning_cvd} construct positive pairs from guarded value-flow paths and AST subtrees of the same semantic fragment, contrasts them against unrelated paths, and employ an attention encoder; an active selection step prioritizes representative or feasible paths, so the contrastive objective aligns precisely with the flow sensitivities on which many vulnerabilities hinge. Generalizing the view design, Zamani et al.~\cite{research_172_cross_domain_vuldetect_graph_contrastive_learning} assemble a cross-domain program graph that unifies AST, control/data-flow, and even binary-level structure, and apply graph augmentations (node/edge perturbations, feature masking) to form views whose agreement teaches invariances that are stable across projects. Complementing contrastive instance discrimination, they adopt a generative objective on graphs with VulMae~\cite{research_173_vulmae}, by masking node features in source/binary control- and property-graphs, and training an encoder–decoder to reconstruct them. This masked-reconstruction pretraining drives the encoder to model local structure and dependencies without labels, after which a light classifier adapts to vulnerability detection. Together, these designs illustrate two self-supervised axes—discriminative alignment (contrastive) and generative reconstruction (masked autoencoding)—and show how program-appropriate “views” (paths, graph augmentations, masked nodes) realize label efficiency in practice.

\textbf{Inspiration.} Similar mechanisms recur across software analytics. Co-training and committee-based querying appear in defect prediction studies~\cite{research_136_sample_based_defect_pred_active_semi_supervised_learning,research_119_effort_aware_semi_supervised_just_in_time_defect_prediction}, where unlabeled modules or commits are pseudo-labeled only under classifier agreement, often coupled with active selection or effort-aware acceptance criteria. Graph-centric semi-supervision surfaces in low-rank and cross-project approaches~\cite{research_231_2018_Low_Rank_semisupervised_defectpred,research_230_2018_Cross_Project_Within_Project_DefectPred_Semisupervised}, which embed unlabeled samples into kernelized, cost-sensitive objectives to regularize under domain shift. PU-style caution appears in log-based anomaly detection~\cite{research_154_semi_supervised_log_based_anomaly_detection_probabilistic_label_estimation}, while mobile security and cyber-physical settings leverage pseudo-labeling and clustering for malware and anomaly detection. These examples suggest transferable strategies for uncertainty calibration, active querying, and hybrid graph-based regularization.

\subsection{RQ1-B: Contrastive Learning}
Contrastive learning shapes an embedding space by pulling positive views closer and pushing negatives apart, typically via InfoNCE or supervised variants. In code, views can be token sequences with augmentation, structure-aware graphs (AST/CFG/DFG/CPG), or cross-modal pairs aligning code and text.

While Section~\ref{subsec41:semi-self} covers semi/self-supervision broadly (including generative self-supervision), we isolate contrastive learning here because it recurs as a reusable objective across token/graph/multimodal settings.

\textbf{Instantiation for CVD.}
Contrastive objectives appear in three main roles: (i) self-supervised pretraining to learn invariances without labels, later adapted with small heads; (ii) supervised variants that compact same-class features and separate classes during fine-tuning; and (iii) domain-aligned formulations combining contrastive geometry with distribution alignment (e.g., MMD or adversarial objectives).

Multi-modal pipelines unify program structure with natural language. CLeVeR~\cite{research_094_clever_multimodal_contrastive_cvd_representation} fuses token-level representations with CPG structure and uses cross-attention to align code with description signals; contrastive pretraining tightens code–description correspondence while preserving structure-sensitive cues. ContraBERT~\cite{research_166_contrabert} augments programs and paired NL, training with momentum contrast and masked LM to yield robust token-space encoders.

Cross-project transfer leverages contrastive geometry plus statistical alignment. Du et al.~\cite{research_257_2023_joint_geometrical_statistical_domain_adapt_vul} combine mutual-nearest-neighbor contrastive objectives with instance reweighting and MMD penalties to reduce domain mismatch. Nguyen et al.~\cite{research_101_CVD_deep_domain_adaptation_max_margin_principle_cross_project} pursue a margin-based alternative, learning domain-invariant features adversarially and imposing max-margin kernels for stable decision regions under label scarcity.

At finer granularity, Nguyen et al.~\cite{research_261_2022_statement_level_vuldetec_learn_vul_patterns_through_info_theory_contrastive_learning} maximize mutual information between statement embeddings and cluster assignments while applying spatially aware contrastive loss; Gumbel-Softmax handles discrete assignments, enabling clustered vulnerability patterns and lightweight classifiers.

Together, these designs illustrate the spectrum of view construction—token augmentations, code–text pairs, mutual-neighbor anchoring across domains, and cluster-aware local objectives—with contrastive learning providing the common mechanism for low-label regimes.

\textbf{Inspiration.}
Supervised contrastive fine-tuning emerges as a simple, effective drop-in on top of pretrained code LMs in defect prediction and related tasks~\cite{research_133_revisiting_semisupervised_unsupervised_methods_effort_aware_cross_project_defect_pred,research_176_effort_aware_jit_defect_pred_unsup_could_be_better_than_sup}. Hybrid or multi-view formulations also appear in vulnerability detection frameworks~\cite{research_268_2025_scalable_vuldetec_system_with_multiview_graph_representation}, while binary analysis tasks exploit contrastive pretraining for universal embeddings~\cite{research_265_2023_generative_denoising_skipgram_unsupervised_binary_code_similarity_detection_gental}. These examples suggest transferable strategies for multi-modal alignment and domain-invariant representation learning.


\subsection{RQ1-C: Few-Shot and Meta-Learning}
Few-shot learning aims to generalize from extremely small labeled sets (e.g., $K$ examples per class) by leveraging prior knowledge across tasks. Meta-learning provides the algorithmic backbone, training models to adapt rapidly to unseen tasks via episodic optimization or metric-based comparisons.

\textbf{Instantiation for CVD.}
Recent systems vary in task framing: function-level classification, clone-style similarity detection, or span localization. Few-VulD~\cite{research_76_2024_Few_VulD_few_shot_learning_framework_for_software_vuldetect} adopts an optimization-based meta-learner in the MAML family, training a lightweight token encoder on program-dependence slices so that a few gradient steps specialize the detector to new vulnerability patterns. Vul-Mirror~\cite{research_78_2020_VulMirror_few_shot_method_for_discovering_vulnerable_code_clone} pursues a metric-based alternative for clone-related vulnerabilities, mapping AST-derived token sequences into an embedding space where similarity scores drive classification. Ju et al.~\cite{research_255_2025_multimodal_framework_for_vuldetec_using_code_simplification_meta_learning} combine multi-modal representation (token encoder + AST GNN) with MAML episodes, aided by code simplification to reduce noise and improve adaptation under imbalance. Corona-Fraga et al.~\cite{research_260_2025_question_answer_methodology_vul_code_review_via_prot_type_maml} extend few-shot learning to question answering and localization via Proto–MAML, optimizing for start/end indices of risky spans while retaining rapid adaptation across CWE types.

\textbf{Inspiration.}
Few-shot/meta-learning also underpins multilingual code understanding. \emph{MetaTPTrans}~\cite{research_258_2023_metatptrans} introduces a language-aware meta-learner that synthesizes encoder parameters conditioned on programming language, enabling specialization under minimal supervision. This design suggests transferable strategies such as task-conditional parameter generation and episodic adaptation for cross-domain vulnerability detection.

\subsection{RQ1-D: Prompt Tuning}
Prompt tuning adapts pretrained code models to downstream vulnerability detection tasks by reformulating classification or localization as a prompted prediction. Unlike manual prompt engineering, which relies on handcrafted templates, prompt tuning learns task-specific prompts—either as discrete tokens (hard prompts), continuous embeddings (soft or prefix prompts), or hybrid forms—while keeping most model parameters frozen. This approach is highly parameter-efficient compared to full fine-tuning.

\textbf{Instantiation for CVD.}
VulPrompt~\cite{research_165_vulprompt} introduces a graph-aware prompt tuning framework where basis prompts condition a GNN classifier over program dependency graphs. Prompts are trained via attention mechanisms that weight their contribution based on graph similarity, while the encoder remains largely frozen. The model is first pretrained on a label-free link prediction task over graphs, injecting structural semantics into node embeddings before prompt-based adaptation.

Wang et al.~\cite{research_157_no_more_finetuning_prompt_tuning_in_code_intelligence} explore token-level prompting with encoder and encoder–decoder models, reformulating vulnerability detection as a cloze-style task using hard templates and verbalizers alongside soft/prefix prompts. Their study finds that prompt tuning consistently outperforms full fine-tuning in low-resource regimes, especially for smaller models, by aligning the downstream objective with the pretraining task (masked language modeling).

Two works enhance the backbone before lightweight adaptation. PDBERT~\cite{research_096_PDBERT} pretrains by predicting control and data dependencies using automatically extracted labels from static analysis, producing a dependency-aware encoder that can later adapt via prompt tuning or simple classifiers. Similarly, StagedVulBERT~\cite{research_097_Stagedvulbert} employs masked statement prediction and a coarse-to-fine architecture to process long code sequences efficiently; after pretraining, the model adapts to function- and statement-level vulnerability detection with minimal labeled data.

\textbf{Inspiration.}
Prompt tuning has been extended to distributed and continual learning scenarios. Studies on federated adaptation apply prompt tuning in multi-modal settings for privacy-preserving training~\cite{research_105_2025_improving_distributed_learning_based_cvd_multi_modal_prompt_tuning}. Continual learning approaches combine hybrid prompts with replay-based and regularization-based techniques to adapt to evolving taxonomies without catastrophic forgetting~\cite{research_106_2024_cvd_prompt_tuning_continual_learning}. Structure-aware prompting has also been explored, where textual prompts are augmented with graph-conditioned prompts derived from code property graphs~\cite{research_74_2025_Structure_Enhanced_Prompt_Learning_Graph_Based_CodeVuldetect}. These examples highlight transferable strategies for federated adaptation, continual learning, and structure-aware prompting in low-resource settings.


\subsection{RQ1-E: LLM-Based Approaches}
\label{subsec45:llm-based}
LLM-based approaches treat source code as natural language and adapt pretrained models either at inference time—via zero-/few-shot prompting, chain-of-thought (CoT) reasoning, and retrieval-augmented generation (RAG)—or at training time through parameter-efficient fine-tuning (e.g., LoRA) and lightweight adapters. These methods aim to exploit the broad prior of code-trained LLMs while minimizing task-specific supervision and engineering overhead.

\textbf{Instantiation for CVD.}
A first family relies on inference-only prompting and retrieval. Zhou et al.~\cite{research_110_llm4cvd_short_results_directions} show that well-crafted few-shot prompts augmented with retrieved CWE exemplars can outperform fine-tuned smaller models, highlighting the importance of prompt design and context provisioning. LLbezpeky~\cite{research_141_leveraging_llms_for_vuldetec} replaces heavy training with a RAG pipeline that selects and summarizes relevant project files for prompt construction, demonstrating that careful context injection can make closed-source models effective without parameter updates. Structured prompting strategies, such as role-based and CoT templates, improve reasoning and vulnerability discrimination~\cite{research_030_2024_Prompt_Enhanced_Software_Vulnerability_Detection_Using_ChatGPT,research_263_2024_enhancing_cvd_llm_prompt_engineering}. Moving to finer granularity, Ceka et al.~\cite{research_269_2024_can_lllm_prompting_serve_as_proxy_for_static_analysis_in_vuldetec} explore CWE-focused instructions and contrastive CoT to teach models to distinguish subtle paired cases within short prompt budgets.

Training-time adaptation complements prompting. Yin et al.~\cite{research_254_2024_multitask_based_evaluation_open_source_llm_on_vul} evaluate decoder-only LLMs under few-shot prompting and LoRA fine-tuning, finding that LoRA significantly improves recall for complex tasks but remains sensitive to input length and domain cues. Kalouptsoglou et al.~\cite{research_275_2025_transfer_learning_svd_using_transformer_models} compare full fine-tuning with embedding extraction for downstream classifiers, while VulDetectBench~\cite{research_252_2024_vuldetectbench} shows that LoRA-based fine-tuning can dramatically raise performance for root-cause localization, yet deep reasoning remains brittle without structural augmentation.

Hybrid pipelines integrate reasoning with external tools. Nong et al.~\cite{research_169_chain_of_thought_prompting_llm_discovering_fixing_sv} parameterize prompts with vulnerability semantics and control/data-flow cues to guide CoT reasoning for detection and patching. Zibaeirad et al.~\cite{research_238_2025_reasoning_llms_zeroshot_vuldetec} enforce structured reasoning templates with confidence checks to improve zero-shot decisions at function-, file-, and multi-file granularity. Li et al.~\cite{research_239_2025_llm_based_vuldetec_were_afraid_to_ask} inject interprocedural context mined from code property graphs into prompts, markedly improving discrimination between patched and vulnerable variants. Beyond detection, PropertyGPT~\cite{research_247_2024_properygpt} and Tihanyi et al.~\cite{research_248_2025_new_era_software_sec_self_healing_llm_formal_verif} close the loop by coupling LLM-generated invariants or patches with formal verification, reducing hallucinations and strengthening correctness guarantees.

\textbf{Inspiration.}
Similar strategies appear in adjacent software engineering tasks. Federated learning approaches integrate prompt tuning and parameter-efficient adaptation for privacy-preserving defect prediction~\cite{research_106_2024_cvd_prompt_tuning_continual_learning}. Neuro-symbolic workflows combine LLM proposers with sound verifiers for automated program verification~\cite{research_248_2025_new_era_software_sec_self_healing_llm_formal_verif}. Multi-modal agents fuse LLM-generated semantic artifacts with graph encodings for smart contract vulnerability detection~\cite{research_235_2025_vuulrag}. These examples highlight transferable strategies for tool-augmented reasoning, federated adaptation, and multi-modal integration in security-critical contexts.

\subsection{RQ1-F: Cross-cutting mechanisms and specialized settings}
These approaches are not stand-alone paradigm families; instead, they are mechanisms or operating
settings that can be composed with multiple families (Sections~\ref{subsec41:semi-self}--\ref{subsec45:llm-based}) to reduce supervision,
improve robustness, or satisfy deployment constraints.

\textbf{Instantiation for CVD.}

\paragraph{Cross-project transfer and domain shift mitigation.}
Here, label efficiency is achieved by reusing labeled knowledge from source projects while controlling negative transfer under shift.
Techniques such as TrAdaBoost and instance reweighting appear repeatedly: CSVD-TF combines handcrafted metrics with CodeBERT embeddings and progressively downweights harmful source samples~\cite{research_104_CSVD_TF}. ALTRA~\cite{research_120_altra_cross_project_active_learning_tradaboost} and DTB~\cite{research_113_negative_samples_cross_company_cvd} integrate instance filtering and active querying to minimize negative transfer under very small target label budgets. Related efforts include unsupervised domain adaptation via MMD alignment~\cite{research_100_CVD_cross_domain_graph_embedding_domain_adaptation} and inspection-oriented objectives such as effort-aware ranking~\cite{research_133_revisiting_semisupervised_unsupervised_methods_effort_aware_cross_project_defect_pred,research_176_effort_aware_jit_defect_pred_unsup_could_be_better_than_sup}.

\paragraph{Efficiency- and scalability-driven designs.}
These works reduce supervision cost \emph{indirectly} by lowering the computation needed to train/adapt models in low-label regimes.
Jianing et al.~\cite{research_063_2025_Enhancing_vulnerability_detection_efficiency_exploration_light_weight_llms_with_hybrid_code_features} apply LoRA-based compression and hybrid AST+CFG fusion to reduce cost without sacrificing recall, while LPASS~\cite{research_244_2025_lpass_linear_probes} uses layer-wise probes to guide pruning and quantization. Complementary lines include heterogeneous graph modeling (HGVul~\cite{research_147_hgvul}) and multi-view graph-to-image conversion~\cite{research_268_2025_scalable_vuldetec_system_with_multiview_graph_representation}, which aim to maintain detection quality while scaling structural processing. Earlier baselines~\cite{research_253_2019_performance_evaluation_vuldetec_methods,research_271_2022_detecting_vul_by_learning_from_large_scale_opensource_repo} likewise highlight the role of slicing, AST paths, and attention mechanisms in stabilizing learning under practical constraints.

\paragraph{Hybrid static/dynamic confirmation pipelines.}
In settings where pure learning-based detection is insufficiently reliable, hybrid pipelines use static or dynamic checks to increase precision with limited labels.
Tevis and Hamilton~\cite{research_273_2006_static_analysis_of_anomalies_security_vulnerabilities_in_exec} inspect PE/COFF metadata to flag structural anomalies without disassembly, while CLORIFI~\cite{research_281_2016_clorifi_cvd_using_clonde_verif} combines syntax-based clone detection with backward sensitive-data tracing and concolic execution to confirm exploitability. Shahmehri et al.~\cite{research_272_2012_advanced_approach_for_detecting_vul} formalize vulnerability conditions as abstract predicates evaluated over instruction traces, avoiding full symbolic exploration.

\paragraph{Specialized settings: binaries and privacy-preserving collaboration.}
Several contributions target operating conditions where labels are scarce due to access, privacy, or platform constraints.
GenTAL~\cite{research_265_2023_generative_denoising_skipgram_unsupervised_binary_code_similarity_detection_gental} learns robust binary embeddings via masked reconstruction, and FedMVA~\cite{research_276_2025_fedmva_federated_multimodal_learning} aggregates tri-modal representations under privacy constraints. Cross-platform binary detection is addressed by VulSeeker~\cite{research_270_2018_vulseeker}, which embeds semantic flow graphs for clone search. Additional niche solutions include predictive scoring from partial CVSS attributes~\cite{research_274_2013_predictive_vul_scoring_in_context_of_insufficient_info_availability} and classical fault-prediction models adapted for vulnerability prioritization~\cite{research_282_2013_traditional_fault_pred_used_for_vulpred}.

\textbf{Inspiration.}
Adjacent software engineering tasks instantiate closely related mechanisms under comparable label constraints.
Li et al.~\cite{research_116_param_optimization_transfer_learning_cross_project_defect_prediction} emphasize that tuning transfer pipelines (filters and feature transforms) can dominate classifier choice in cross-project settings. Human-in-the-loop approaches such as EMBLEM~\cite{research_151_better_data_labeling_with_emblem} optimize labeling cost for commit-level defect prediction. Studies such as CLAMI~\cite{research_232_2015_unsupervised_defectpred_clami} explore clustering and active learning for label-efficient defect prediction, while Smart-Cuts~\cite{research_102_active_learining_CVD_pruning_bad_seeds} improves acquisition stability by filtering harmful seeds. Unsupervised domain adaptation with metric alignment further supports heterogeneous prediction without target labels~\cite{research_155_unsupervised_deep_domain_heterogenous_defect_pred}. These examples illustrate transferable design choices for cross-project robustness and label acquisition beyond CVD.

\begin{rqsummary}{RQ1 — Summary}
\begin{enumerate}
    \item Label-efficient CVD methods span semi-/self-supervision, contrastive pretraining, few-shot/meta-learning, prompt tuning, LLM-based adaptation, and cross-project transfer.
    \item Semi-/self-supervised and contrastive methods dominate among CVD-specific studies, while active learning and anomaly detection appear more in adjacent SE tasks.
    \item Most paradigms exploit either unlabeled code (self-/semi-supervision), structural in-variances (contrastive), or rapid adaptation (few-shot and prompting).
\end{enumerate}
\end{rqsummary}

\section{RQ2: Design Choices}
\label{sec:rq2}

RQ2 characterizes how label-efficient CVD studies are operationalized in practice: the representation used to encode code artifacts, the task framing (binary detection vs.\ CWE typing), and the evaluation protocols that determine comparability across papers. Beyond reporting frequencies, this section highlights how these design choices constrain what current evidence supports and where conclusions may fail to transfer across datasets, languages, and deployment settings.

\subsection{RQ2-A: Code Representation}
Representation choices shape what information a label-efficient CVD pipeline can exploit under limited supervision: lexical cues, structural semantics, or external signals from analysis tools. They also determine practical costs (graph construction, normalization, context limits) and, by extension, which paradigms are feasible at scale. In the Main pool, we observe four recurring representation families: token-based, graph-based, hybrid (multi-view, typically fusing tokens and graphs), and knowledge-based (hand-engineered signals derived from program analysis and software engineering metadata, e.g., static-analysis warnings, complexity/maintainability metrics, dependency or change-history features).

\begin{figure}[h]
  \centering
  \begin{subfigure}[t]{0.5\textwidth}
    \centering
    \includegraphics[width=\linewidth]{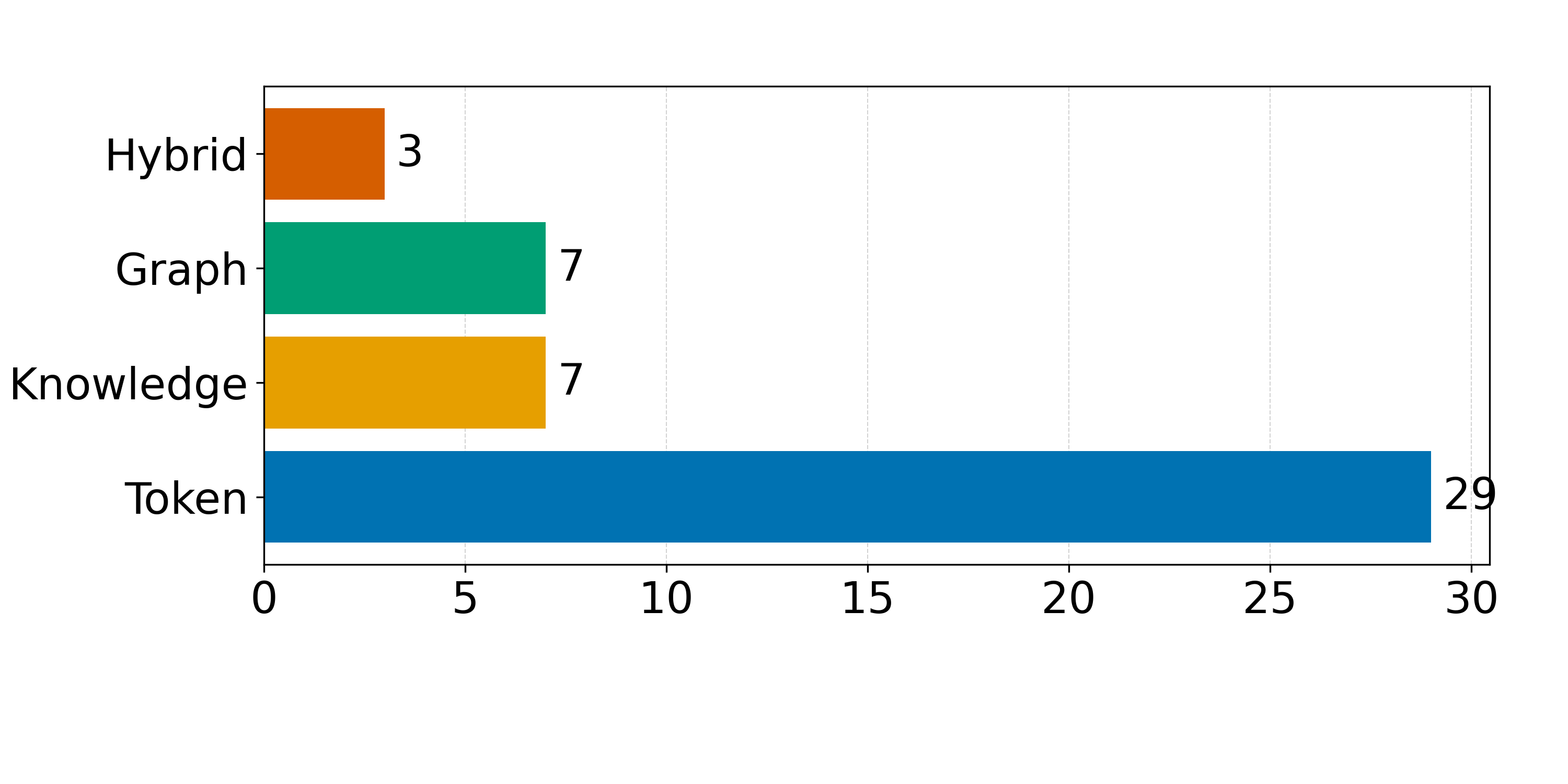}
    \caption{Representation distribution by family}
    \label{fig:representation_family}
  \end{subfigure}\hfill
  \begin{subfigure}[t]{0.5\textwidth}
    \centering
    \includegraphics[width=\linewidth]{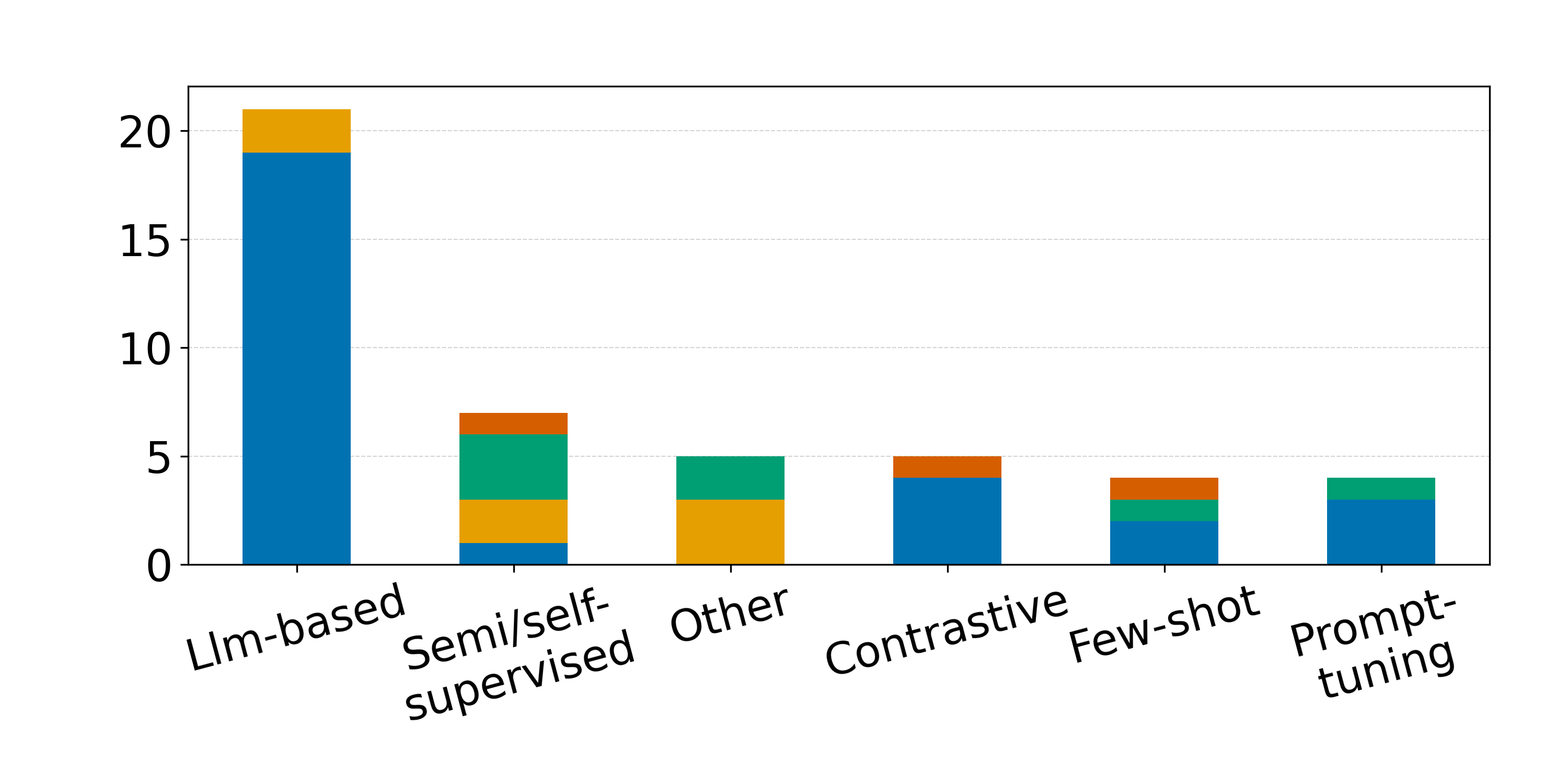}
    \caption{Paradigm $\times$ family counts}
    \label{fig:representation_familyxparadigm}
  \end{subfigure}
  \caption{Code representation trends in the Main pool.}
  \Description{Two bar charts: (a) Token dominates with 29 studies; graph and knowledge-based have 7 each; hybrid has 3. (b) Paradigm $\times$ family counts, with token-heavy LLM-based pipelines and more frequent graph/hybrid use in semi/self-supervised and contrastive settings.}
  \label{fig:representation_trends}
\end{figure}

\textbf{Families and prevalence.}
Token-based representations dominate (29 studies), reflecting the widespread use of pretrained code language models and parameter-efficient adaptation. Graph-based and knowledge-based representations each appear in seven studies, while explicit hybrid designs remain comparatively rare (three studies), suggesting that multi-view pipelines are still emerging rather than standard practice (Figure~\ref{fig:representation_trends}).

\textbf{Paradigm alignment.}
Figure~\ref{fig:representation_familyxparadigm} indicates that LLM-based pipelines are overwhelmingly token-centric, consistent with their reliance on sequence encoders and prompt-oriented interfaces. In contrast, non-LLM label-efficiency mechanisms more often incorporate structure or external signals: graph-based views and hybrids appear more frequently in semi/self-supervised and contrastive settings, where relational context and invariances can provide supervision when labels are scarce. This association should be read as an empirical tendency in the surveyed literature (not a strict requirement): token-only methods can be used beyond LLM pipelines, and graph-based methods can be integrated with pretrained encoders, but the Main pool shows clear clustering.

\begin{figure}[h]
\centering
\includegraphics[height=0.5\linewidth, width=0.6\linewidth]{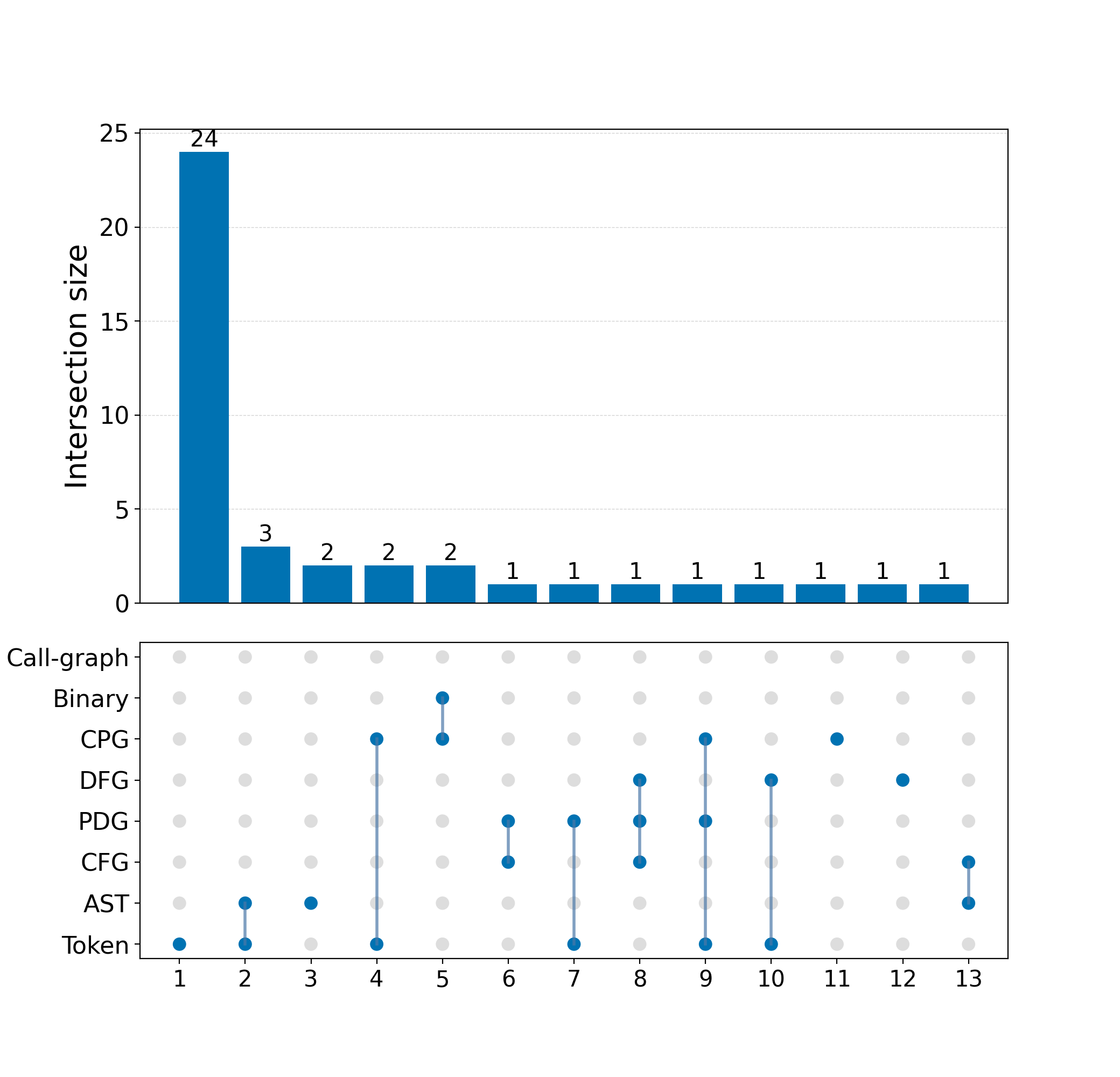}
\caption{Primitive-level intersections across representations.}
\Description{Upset plot of primitive co-occurrence across surveyed papers. Token-only designs form the largest intersection (size 24). Smaller intersections combine AST and control/data-flow primitives (e.g., CFG/PDG/CPG), indicating limited but present multi-primitive usage.}
\label{fig:representation_primitives}
\end{figure}

\textbf{Representation primitives and co-occurrence.}
Beyond broad families, Figure~\ref{fig:representation_primitives} details which primitives are used and combined. The strongest single intersection corresponds to token-only designs (intersection size 24), highlighting their simplicity and compatibility with pretrained encoders. Multi-primitive designs occur but remain fragmented: when graphs are used, they typically involve AST and/or control/data-flow variants (e.g., CFG/PDG/CPG), and truly multi-view combinations are uncommon. This matters because different primitives encode different inductive biases: AST emphasizes syntactic structure, CFG/PDG capture control/data dependencies, and CPG integrates multiple relations. As a result, comparing “graph-based” approaches without specifying primitives can obscure why a method works (or fails) under label scarcity.

\textbf{Implications.}
Token-based representations are currently the default for scalable evaluation and rapid adaptation, but their dominance also narrows what is empirically validated about structure-aware supervision saving. Graph-enhanced and hybrid pipelines may offer stronger semantic inductive bias, yet they introduce additional degrees of freedom (graph extraction choices, normalization, granularity alignment) that complicate reproducibility and fair comparison.

\subsection{RQ2-B: Task Type}
Task type specifies the prediction target in code vulnerability detection. In the Main pool, studies overwhelmingly formulate CVD as \emph{binary classification} (vulnerable vs.\ non-vulnerable), while a smaller subset targets \emph{CWE typing} (multi-class classification). This choice is not merely a modeling detail: it changes the annotation burden, the difficulty profile (long-tailed labels), and what downstream actions the model can support.

\begin{figure}[h]
    \centering
    \begin{subfigure}[b]{0.49\textwidth}
        \includegraphics[width=\textwidth]{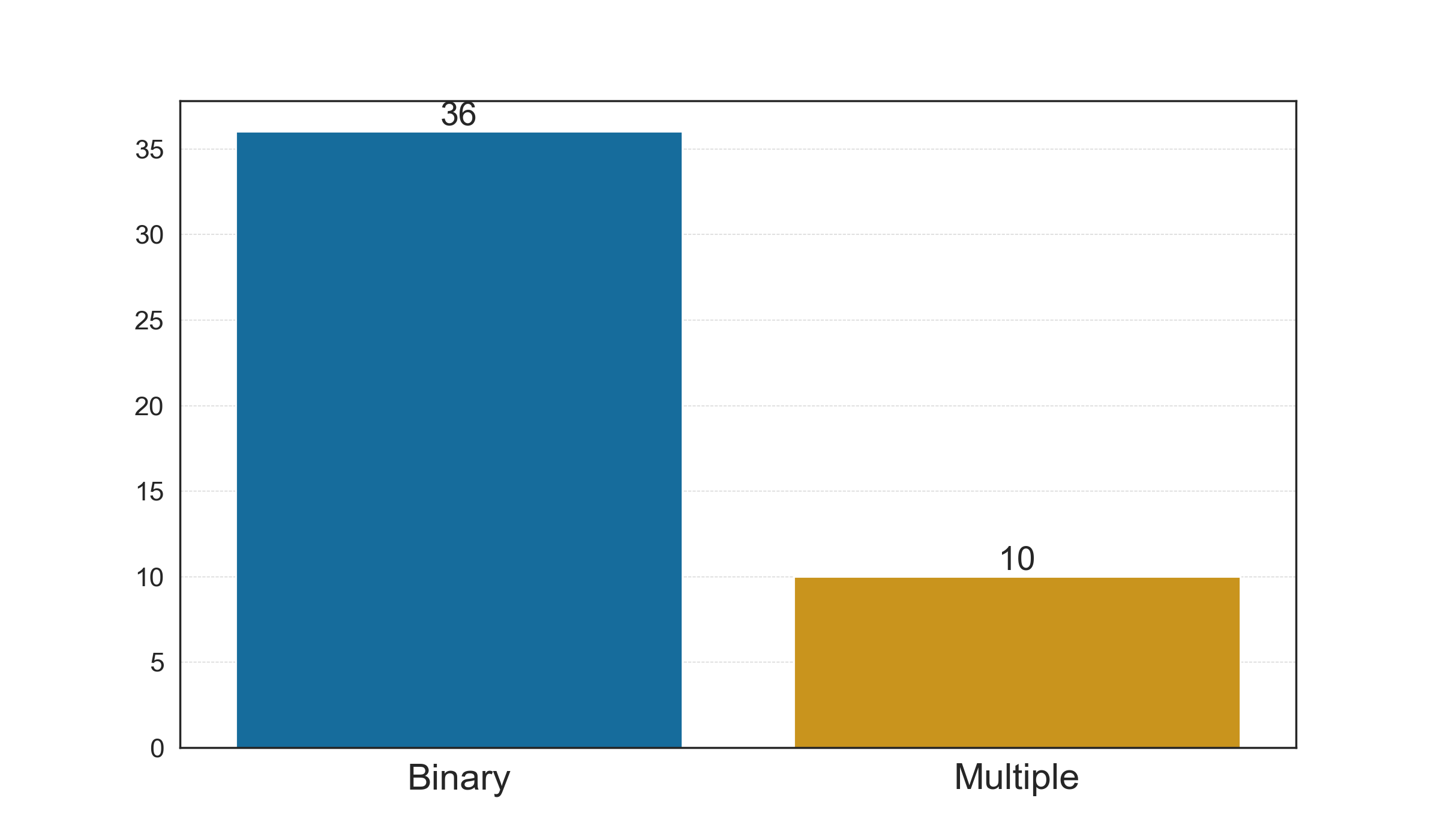}
        \caption{Task type distribution}
        \label{fig:task-type-bin}
    \end{subfigure}
    \hfill
    \begin{subfigure}[b]{0.49\textwidth}
        \includegraphics[width=\textwidth]{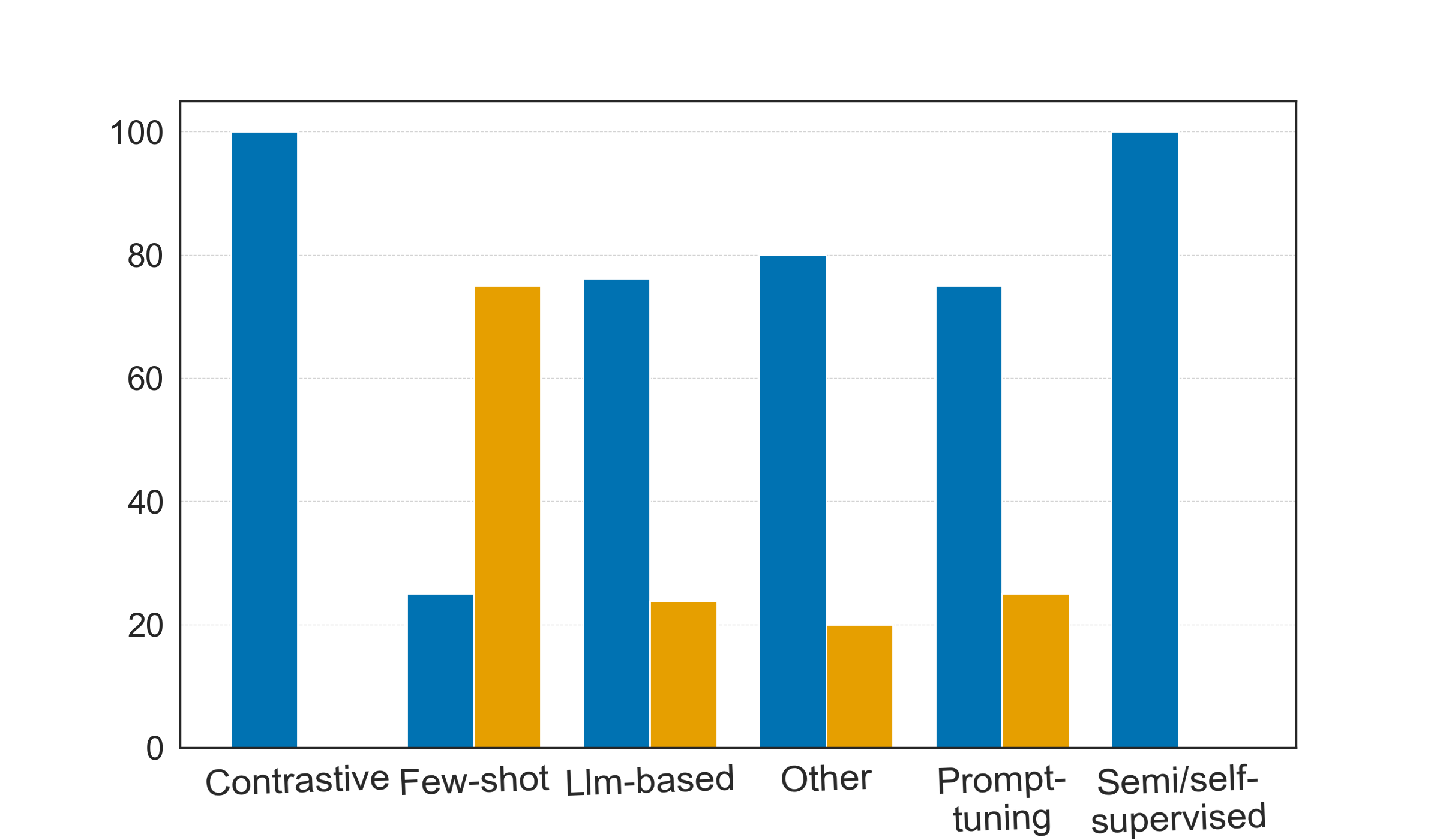}
        \caption{Task type per paradigm}
        \label{fig:task-type-grouped}
    \end{subfigure}
    \caption{Distribution of task types and their association with learning paradigms.}
    \Description{Two bar charts summarizing task type trends in the Main pool. Panel (a) compares binary classification (36 studies) against multi-class CWE classification (10 studies). Panel (b) breaks down task type by paradigm: binary tasks dominate most paradigm families, while multi-class tasks appear less frequently and cluster in a smaller subset of approaches.}
    \label{fig:task-type}
\end{figure}

Figure~\ref{fig:task-type-bin} shows that binary detection dominates (36 studies), whereas CWE classification is comparatively less common (10 studies). A practical driver is supervision: CWE typing requires reliable fine-grained labels and typically exhibits severe class imbalance, making it harder to evaluate and to scale under low-label regimes.

\textbf{Association with learning paradigms.}
Figure~\ref{fig:task-type-grouped} suggests that binary framing is used across all paradigm families, while CWE typing concentrates in a smaller portion of the literature (mainly few-shot and meta-learning). This aligns with how label-efficiency mechanisms are commonly instantiated: binary settings integrate naturally with pseudo-labeling, consistency regularization, and contrastive objectives, whereas CWE typing often requires additional structure (e.g., episodic formulations, label semantics, or prompt scaffolding) to remain tractable under sparse supervision. Importantly, this distribution limits what the current evidence supports: many ``label-efficient'' gains are demonstrated primarily for binary screening, while the literature provides fewer results on fine-grained vulnerability categorization.

\textbf{Implications.}
Because binary and CWE-typing tasks stress different failure modes (false-alarm burden vs.\ long-tailed confusion among weakness classes), results should be interpreted within their task framing. When studies claim general improvements in label efficiency, reporting should clarify whether gains transfer beyond binary detection and under what label distributions. Expanding CWE-aware benchmarks and evaluation under long-tailed settings is necessary to assess whether representation choices and supervision-saving mechanisms generalize beyond screening-oriented formulations.

\subsection{RQ2-C: Evaluation Metrics}
Evaluation metrics determine how label-efficient CVD pipelines are interpreted and compared. This is particularly consequential in CVD, where datasets are typically noisy and strongly imbalanced: a single headline score can hide large differences in false-alarm burden, missed-vulnerability risk, and thresholding choices. Overall, reported evaluation remains concentrated around a small set of classification metrics, with limited diversity in how operational trade-offs are characterized.

\begin{figure}[h]
    \centering
    \includegraphics[width=0.5\linewidth]{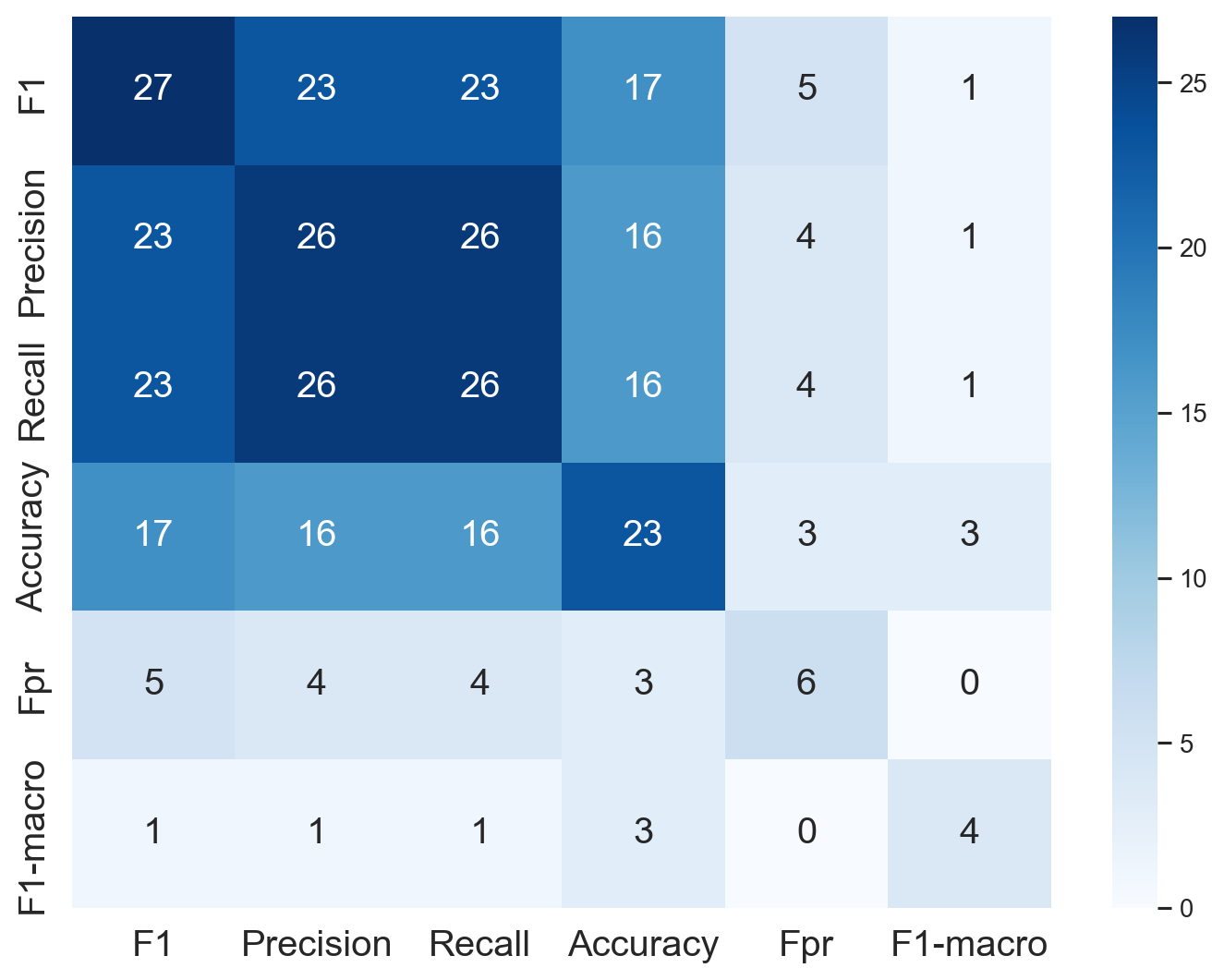}
    \caption{Heatmap of co-reporting among top metrics.}
    \Description{Matrix heatmap of metric co-reporting across surveyed papers. F1, precision, and recall are frequently reported together, whereas other metrics show substantially lower co-reporting.}
    \label{fig:heatmap-metrics}
\end{figure}

\textbf{Observed reporting practice (co-reporting).}
Figure~\ref{fig:heatmap-metrics} summarizes which metrics are reported together across the surveyed studies. Co-reporting concentrates on \emph{F1, precision, and recall}, which frequently appear as a bundle, while other metrics are comparatively less common. This pattern suggests that many papers evaluate at a single operating point using closely related measures, rather than triangulating performance with complementary perspectives (e.g., threshold-robustness, imbalance robustness, or deployment cost).

Heavy reliance on F1/precision/recall alone could make cross-paper comparisons fragile for three recurring reasons. First, these metrics are \emph{threshold-dependent}: two models with similar F1 can imply very different triage workloads depending on how the threshold is chosen. Second, they do not directly expose \emph{prevalence sensitivity} under severe imbalance; improvements can reflect favorable class ratios or sampling rather than robust vulnerability discrimination. Third, they are not explicitly \emph{effort-aware}: security workflows often prioritize reducing analyst time and accelerating discovery, so ranking/inspection effectiveness can matter more than a marginal gain in pointwise F1.

To make results comparable and practically meaningful without increasing evaluation overhead substantially, we recommend that studies reporting F1/precision/recall additionally include:
(i) at least one \emph{imbalance-robust} companion metric (e.g., MCC or PR-AUC when available),
(ii) a clear statement of the \emph{operating point} (how thresholds are selected and knowable at test time),
and (iii) a \emph{label-efficiency curve} (performance vs.\ label budget) rather than a single-point estimate.
Where the intended use is triage, adding an effort-oriented view (e.g., top-$k$ inspection effectiveness or cost-weighted errors) better reflects deployment constraints than optimizing F1 alone.

\subsection{RQ2-D: Datasets and Languages}
Dataset and language choices largely determine what “works” in CVD, because they fix the underlying vulnerability distribution, labeling process, and the kinds of confounders a model can exploit (project or style-specific cues, duplicate code, or patch-related artifacts). In practice, the Main pool relies on a small set of recurring benchmarks and a narrow language slice, which strengthens comparability within that slice but limits external validity.

\begin{figure}[h]
    \centering
    \begin{subfigure}[b]{0.49\textwidth}
        \includegraphics[width=\textwidth]{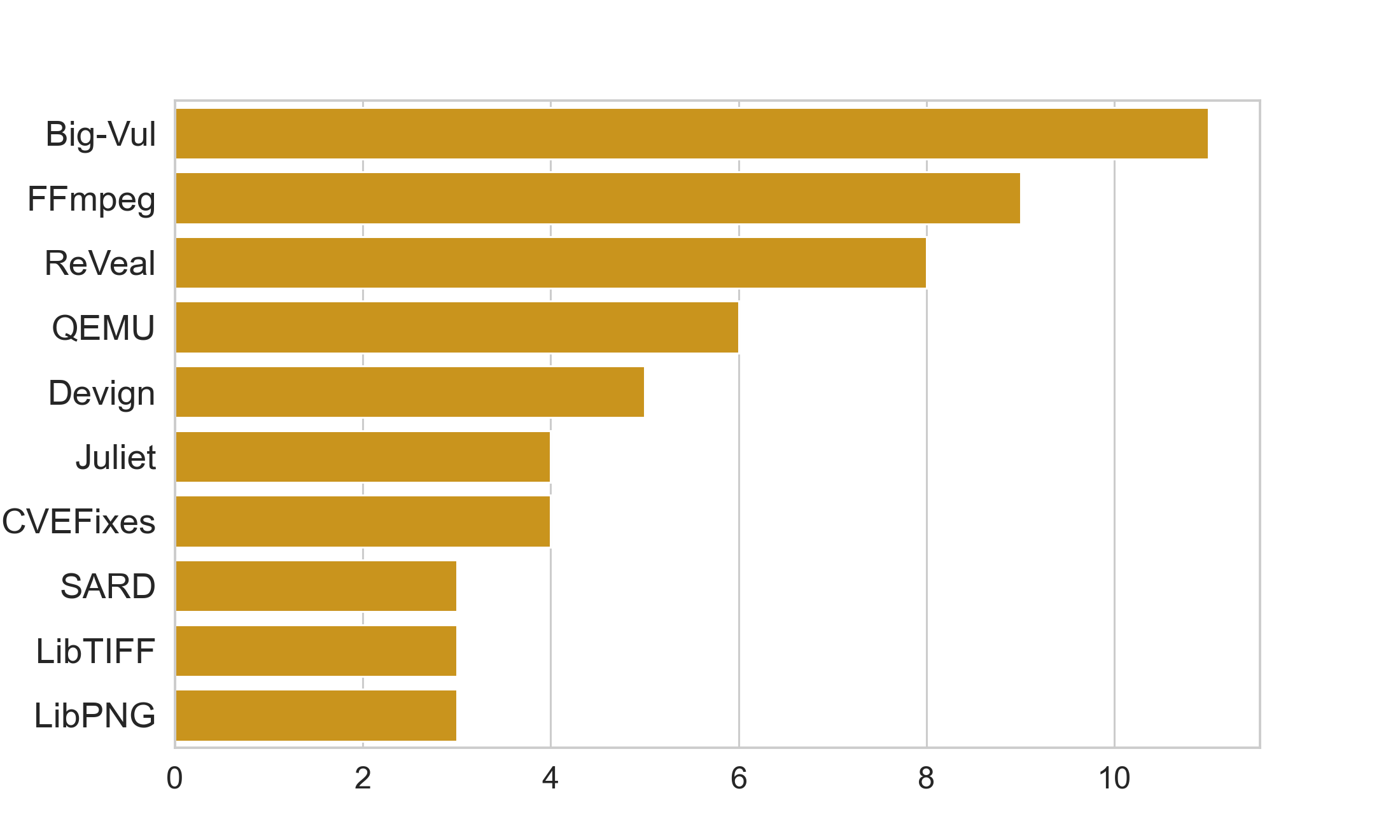}
        \caption{Top datasets used}
        \label{fig:datasets}
    \end{subfigure}
    \hfill
    \begin{subfigure}[b]{0.49\textwidth}
        \includegraphics[width=\textwidth]{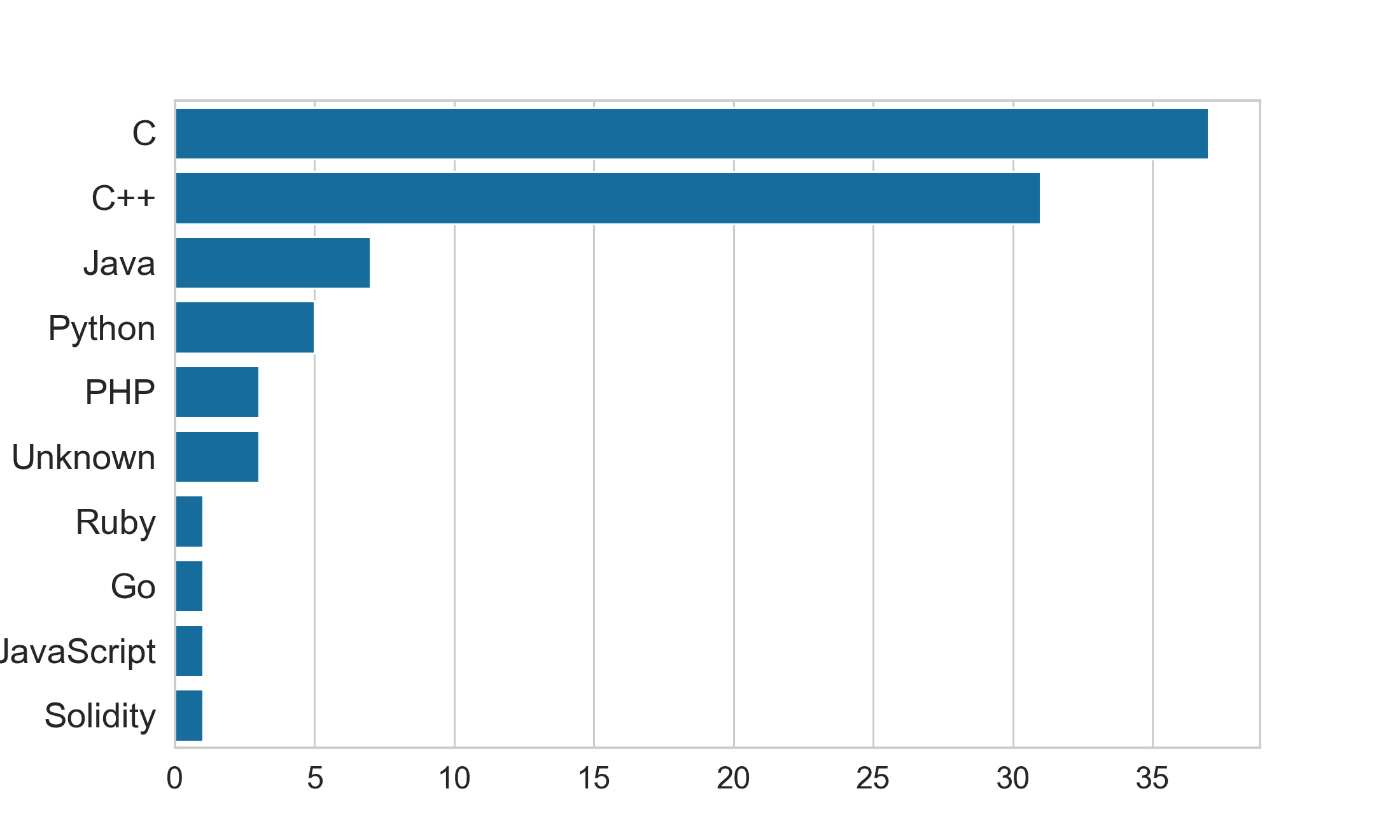}
        \caption{Top programming languages}
        \label{fig:languages}
    \end{subfigure}
    \caption{Dataset and language distribution in the Main pool.}
    \Description{Two bar charts. (a) Big-Vul is the most used dataset, followed by FFmpeg, ReVeal, QEMU, Devign, Juliet, CVEFixes, SARD, LibTIFF, and LibPNG. (b) Most studies focus on C and C++, followed by smaller representation of Java, Python, PHP, and minimal presence of Ruby, Go, JavaScript, and Solidity.}
    \label{fig:datasets-languages}
\end{figure}

\textbf{Dataset concentration.}
Figure~\ref{fig:datasets} shows that a few corpora dominate experimental evaluation, with Big-Vul appearing most frequently, followed by project-specific datasets (e.g., FFmpeg, QEMU) and established curated sets (e.g., ReVeal, Devign). This concentration has two consequences. First, repeated reuse of the same benchmarks can inflate perceived progress if models overfit dataset-specific regularities rather than vulnerability semantics. Second, project-centric corpora risk conflating “vulnerable vs.\ non-vulnerable” with “this project’s coding style, API usage, or commit conventions,” which can be especially problematic when claims target cross-project deployment.

\textbf{Language coverage.}
Figure~\ref{fig:languages} confirms a strong bias toward C and C++, consistent with their prevalence in systems software and memory-unsafe vulnerability classes. However, this focus also means that many conclusions implicitly reflect C/C++-specific signals (e.g., pointer-heavy idioms, macro usage, low-level API patterns) and may not transfer to managed or scripting languages. Java and Python appear only sporadically, while PHP, Ruby, Go, and Solidity are rarely represented, leaving multilingual and heterogeneous-codebase settings underexplored.

Given the dataset and language concentration, strong in-dataset results should not be read as evidence of broad generalization. At minimum, studies should (i) document split protocols and leakage controls (e.g., de-duplication and project-/time-aware splits when applicable), (ii) report dataset characteristics that affect difficulty (size, label source, CWE distribution/coverage), and (iii) include at least one robustness check (cross-project, cross-version, or cross-dataset evaluation) when making deployment-oriented claims. Broadening evaluation beyond C/C++ and beyond a handful of benchmarks is necessary to establish whether representation choices and label-efficiency mechanisms transfer across domains.

\subsection{RQ2-E: Granularity and Availability}
Granularity determines what a vulnerability detector is asked to do: coarse screening (project/file) versus actionable localization (function/line/statement). This choice shapes the modeling pipeline (what context is visible), the evaluation protocol (what constitutes a correct prediction), and the operational usefulness of the output. Artifact availability (code, splits, preprocessing, and evaluation scripts) then determines whether reported gains can be verified and compared.

\textbf{Granularity concentration.}
Table~\ref{tab:granularity} shows that the Main pool overwhelmingly evaluates at the function level (28 studies). Finer-grained settings (line/statement) appear only in limited instances, and multi-granular designs are uncommon. This concentration makes benchmarking easier, but it also narrows what current evidence supports: a function-level detector can rank functions for inspection, yet it does not directly answer \emph{where} the vulnerability is or how much code an analyst must review.

\begin{table}[!htbp]
\centering
\vspace{-4mm}
\caption{Granularity combinations of vulnerability detection tasks in the Main pool (intersection counts).}
\label{tab:granularity}
\scriptsize
\begin{tabular}{c c}
\toprule
\textbf{Granularity combination} & \textbf{\# studies} \\
\midrule
Function-level only & 28 \\
Function-level + statement-level & 3 \\
Function-level + line-level & 3 \\
Project-level only & 3 \\
Statement-level only & 2 \\
Line-level only & 2 \\
Function-level + file-level + multi-file & 2 \\
Project-level + file-level & 1 \\
Function-level + file-level & 1 \\
File-level only & 1 \\
\bottomrule
\end{tabular}
\vspace{-4mm}
\end{table}

Although many papers rely on public datasets, reproducibility is often limited by missing implementation details and incomplete releases. In particular, the parts most likely to affect outcomes; preprocessing and de-duplication, graph construction and normalization, prompting templates and decoding settings, and the exact train/validation/test splits; are not consistently provided. As a result, it is difficult to attribute differences in performance to the proposed label-efficiency mechanism rather than to pipeline-specific choices.

The dominance of function-level benchmarks risks over-optimizing for screening-only use cases and slowing progress on localization-oriented settings that better match developer workflows. To strengthen comparability without imposing large overhead, we recommend that studies (i) explicitly justify their granularity with respect to the intended use case (screening vs.\ localization), and (ii) provide a minimal replication package: fixed splits (with leakage controls), preprocessing scripts, model configuration (including prompting/decoding where relevant), and evaluation code. When new granularities are introduced, reporting should include effort-aware measures (e.g., top-$k$ inspection effectiveness) to connect predictions to practical triage.

\begin{rqsummary}{RQ2 — Summary}
\begin{enumerate}
    \item \textbf{Representation:} Token-based encodings dominate, particularly in LLM and prompting pipelines; graph-based and hybrid (token+graph) representations appear less frequently and are more common in structure-aware settings (e.g., semi/self-supervised and contrastive pipelines).
    \item \textbf{Task framing:} Binary vulnerability detection is far more common than CWE typing, reflecting the higher annotation burden and long-tailed label distributions in fine-grained settings.
    \item \textbf{Metrics:} Reporting concentrates on F1/precision/recall; complementary imbalance-robust and effort-aware metrics remain comparatively underused.
    \item \textbf{Benchmarks:} Evaluation relies on a small set of recurring datasets and is heavily skewed toward C/C++ code; multilingual and cross-domain benchmarks are uncommon.
    \item \textbf{Granularity \& artifacts:} Function-level evaluation dominates; reproducibility is uneven due to incomplete artifact releases and missing preprocessing/splitting details.
\end{enumerate}
\end{rqsummary}

\section{RQ3: Label budgets}
\label{sec:rq3}

We analyze 46 CVD studies in the Main Set. Label budgets are not reported in a standardized way: studies mix different units
(K-shot, labeled fractions, absolute counts, or implicit supervision), and a substantial portion either omits budgets or operates
in zero-shot/no-training regimes. Based on our extracted data, we observe three dominant budget-reporting patterns, alongside a large
``unspecified'' remainder (not reported/unclear).

\subsubsection{Budget specification patterns in CVD}

\subsubsection{K-shot (few-shot) budgets}
A first pattern specifies supervision through \emph{K-shot} support sets, i.e., budgets expressed as a small number of labeled examples
per type/pattern. In our Main Set, 8/46 studies (17.4\%) use a K-shot budget. This reporting style is common in fine-grained settings
such as CWE-typed learning or vulnerability matching, as in \emph{Few-VulD}~\cite{research_76_2024_Few_VulD_few_shot_learning_framework_for_software_vuldetect}.
These studies typically report shots (rather than dataset-level labeled percentages), which can make annotation effort hard to compare against fraction-based regimes.

\subsubsection{Fraction-based budgets (percentage labeled)}
A second pattern specifies supervision as a labeled fraction (e.g., 10\%, 20\%). In our Main Set, 6/46 studies (13.0\%) report budgets
primarily as percentages. This style is characteristic of semi-supervised and PU pipelines where a small labeled subset seeds training,
including teacher-student self-training with uncertainty filtering~\cite{research_74_2025_less_is_more_semi_supervised_deep_learning_vuldetec},
mixed-supervision PU training~\cite{research_73_2023_less_is_enough_positive_unlabeled_model_for_vuldetec}, and graph-based consistency
methods~\cite{research_124_path_sensitive_embedding_contrastive_learning_cvd}. In these papers, label-efficiency claims are commonly tied
to filtering/selection mechanisms (e.g., confidence thresholds and consistency constraints).

\subsubsection{Qualitative or implicit budgets (including LLM/anomaly-style settings)}
A third pattern describes label scarcity qualitatively (``scarce labels'', ``large unlabeled pool'') or treats supervision as implicit
(via exemplars/demonstrations/expert feedback) rather than as an explicit labeled budget. In our Main Set, 11/46 studies (23.9\%) fall
into a zero-shot/no-training regime, and these are concentrated in LLM-based paradigms (9 such cases in the LLM-based subset).
Representative CVD examples include anomaly-style vulnerability identification such as \emph{ANVIL}~\cite{research_088_2024_Anvil} and
expert-in-the-loop prompting with GPT-4o~\cite{research_162_expert_in_the_loop_cross_domain_in_domain_fsl_svd}. In these settings, papers
rarely report comparable budget variables (e.g., number of exemplars, number of expert decisions, or context/token constraints), limiting
comparability with K-shot and percentage-based reporting.

\subsubsection{Limitations in label-budget reporting}
Label-efficiency evidence is reported with heterogeneous conventions. Studies specify supervision using different units
(K-shot support sets, labeled fractions, absolute labeled counts, or implicit supervision in zero-shot/no-training settings). In particular, multi-budget evaluations (e.g., varying the labeled fraction
or the number of shots) remain less consistently reported than single-budget results, and implicit-supervision settings (e.g., LLM prompting)
rarely expose budget variables that are comparable to labeled data (such as the number of exemplars, retrieval depth, or context constraints).
These observations motivate clearer, more comparable budget descriptions in future CVD work.

\begin{rqsummary}{RQ3 — Summary}
\begin{enumerate}
    \item Label budgets in CVD are specified using heterogeneous units, ranging from K-shot support sets and labeled fractions to absolute counts and implicit supervision in zero-shot/no-training settings.
    \item Three dominant reporting patterns emerge: (i) K-shot budgets for fine-grained settings, (ii) percentage-labeled budgets for semi-supervised pipelines, and (iii) qualitative or implicit budgets for LLM-style approaches.
    \item K-shot reporting emphasizes support-set size rather than dataset-level annotation effort, while fraction-based reporting is typically tied to filtering/selection mechanisms that leverage unlabeled data.
 
\end{enumerate}
\end{rqsummary}


\section{RQ4: Computing Cost}
\label{sec:rq4}

\paragraph{Coding frame.}
We treat \emph{compute cost} as any information that characterizes (i) preprocessing overhead, (ii) training/adaptation overhead,
or (iii) inference/serving overhead. We separately record \emph{compute resources} (e.g., CPU/GPU/RAM, node counts, environments)
and whether studies mention \emph{API-based} inference (hosted LLMs) or provide \emph{explicit monetary} cost analysis.

\paragraph{How often is cost reported?}
Across the Main Pool, compute reporting is partial: $28/46$ studies provide some compute-cost signal (e.g., time,
epochs, number of calls), while $34/46$ report compute resources (hardware/environment). Only $22/46$ report both, and $6/46$ report neither.
In other words, cost is \emph{not consistently} documented, and most reporting emphasizes \emph{machines} rather than end-to-end operational cost.

\paragraph{Is label-efficiency ``expensive''? It depends on where the cost moves.}
We observe a consistent \emph{cost-shifting} pattern rather than uniformly ``cheap'' label-efficiency:
(i) \textbf{Graph/hybrid pipelines} often trade labels for heavier preprocessing/engineering (e.g., slicing/graph construction dominating runtime~\cite{research_146_web_app_vul_pred_hybrid_program_analysis_ml,research_096_PDBERT}),
(ii) \textbf{semi-/self-supervised and contrastive training} may reduce labels but introduce additional training complexity (extra objectives,
negative sampling, longer training schedules~\cite{research_124_path_sensitive_embedding_contrastive_learning_cvd,research_74_2025_less_is_more_semi_supervised_deep_learning_vuldetec}),
and (iii) \textbf{LLM-based pipelines} frequently move the dominant cost to inference time (context length, retrieval, multiple passes~\cite{research_031_2023_How_Far_Have_We_Gone_in_Vulnerability_Detection_Using_LLMs,research_088_2024_Anvil,research_110_llm4cvd_short_results_directions}),
sometimes under API rate/latency constraints.

\paragraph{Beyond hardware: do studies discuss API usage and monetary cost?}
API/hosted-LLM usage is mentioned in $14/46$ studies, but explicit monetary analysis is rare~\cite{research_141_leveraging_llms_for_vuldetec,research_255_2025_multimodal_framework_for_vuldetec_using_code_simplification_meta_learning,research_269_2024_can_lllm_prompting_serve_as_proxy_for_static_analysis_in_vuldetec}. When present, it typically appears as
(i) limits on the number of evaluated instances due to API cost, (ii) counts of API calls/tokens, or (iii) a small token-cost analysis.
This suggests that \emph{hardware specs alone} are an incomplete proxy for operational expense in LLM-centric CVD, and that the community
still under-reports cost drivers that matter for deployment (latency, throughput, context window policies, and API billing).

\begin{rqsummary}{RQ4 — Summary}
\begin{enumerate}
    \item Label-efficient CVD is \emph{not} uniformly ``cheap''; instead, studies often reduce labels by shifting burden to preprocessing,
    training complexity, or inference-time compute.
    \item At the same time, cost reporting remains inconsistent and skewed toward
    hardware descriptions, with limited explicit accounting of API-driven expenses.
 
\end{enumerate}
\end{rqsummary}


\section{RQ5: Reported limitations}
\label{sec:rq5}

We synthesize two complementary types of limitations in the Main pool: (i) \emph{limitations explicitly stated by authors} (e.g., narrow benchmarks, external validity, operational constraints, mechanism brittleness), and (ii) \emph{recurring reporting omissions observed during our extraction} that hinder verification, replication, and cross-paper comparison.

Across the corpus, author-stated limitations most frequently concern (i) narrow dataset/language coverage and labeling artifacts,
(ii) limited external validity (cross-project/language robustness), (iii) operational constraints (compute/latency/API usage), and
(iv) mechanism-specific brittleness (e.g., augmentation choice, support-set sensitivity, prompt/context sensitivity). In addition,
we observe recurring \emph{reporting omissions}; notably around data protocols, splits/deduplication, compute, and inference settings; that motivate the checklist in the final subsection.

\subsection{Data, labeling, and benchmark limitations}
A substantial fraction of studies explicitly note that conclusions are drawn from narrow evaluation settings; often a small
number of projects, a single benchmark family, or a single language (e.g., PHP-only web settings~\cite{research_146_web_app_vul_pred_hybrid_program_analysis_ml} or C/C++-only corpora~\cite{research_74_2025_less_is_more_semi_supervised_deep_learning_vuldetec}).

Label quality and construct validity are also recurring concerns: several works mention residual dataset noise, weak/heuristic
labels, or evaluation artifacts that can inflate apparent gains~\cite{research_73_2023_less_is_enough_positive_unlabeled_model_for_vuldetec,research_238_2025_reasoning_llms_zeroshot_vuldetec}. For LLM-based pipelines, some papers additionally raise the risk of implicit leakage or
confounding when relying on closed-source systems or LLM-as-judge evaluation \cite{research_110_llm4cvd_short_results_directions}.

\subsection{Generalization and external validity}
Many approaches are evaluated in-distribution and may not transfer reliably across projects, domains, or languages~\cite{research_172_cross_domain_vuldetect_graph_contrastive_learning,research_157_no_more_finetuning_prompt_tuning_in_code_intelligence}. More broadly, papers report performance drops under distribution shift, differences in coding style, or changes in vulnerability
types, which complicate claims about practical deployability.

\subsection{Operational constraints: compute, latency, and hosted-LLM usage}
Operational constraints are repeatedly cited as barriers to adoption and reproducibility, spanning (i) heavy preprocessing/toolchain
requirements (e.g., graph extraction), (ii) training/adaptation overhead, and (iii) inference/serving constraints. For example,
graph and LLM-assisted pipelines may depend on external analyzers or retrieval components, and LLM-based approaches highlight token
budgets, context-window truncation, rate limits, and cost considerations~\cite{research_169_chain_of_thought_prompting_llm_discovering_fixing_sv,research_235_2025_vuulrag}.

\subsection{Paradigm-specific failure modes (mechanism-linked)}
Beyond cross-cutting issues, several failure modes recur in ways that align with the underlying label-efficient mechanisms:

\paragraph{Semi-/self-supervised.}
Methods that leverage unlabeled data can be sensitive to dataset noise and evaluation artifacts, and improvements may not persist
outside the original benchmark family~\cite{research_142_predicting_buffer_overflow_using_semi_supervised_learning,research_73_2023_less_is_enough_positive_unlabeled_model_for_vuldetec}. 

\paragraph{Contrastive learning and augmentation design.}
Contrastive objectives may depend strongly on the definition of views/augmentations and negative sampling, with some studies noting
that gains hinge on these design choices or are uneven across tasks~\cite{research_166_contrabert,research_094_clever_multimodal_contrastive_cvd_representation}.

\paragraph{Few-shot / expert-in-the-loop regimes.}
Few-shot pipelines can be brittle with respect to support-set composition, data scarcity, and false positives~\cite{research_76_2024_Few_VulD_few_shot_learning_framework_for_software_vuldetect,research_78_2020_VulMirror_few_shot_method_for_discovering_vulnerable_code_clone,research_162_expert_in_the_loop_cross_domain_in_domain_fsl_svd}. 

\paragraph{LLM prompting and context sensitivity.}
LLM-based approaches frequently report sensitivity to prompts, insufficient context due to token limits, and instability across
decoding settings or datasets~\cite{research_169_chain_of_thought_prompting_llm_discovering_fixing_sv,research_031_2023_How_Far_Have_We_Gone_in_Vulnerability_Detection_Using_LLMs,research_239_2025_llm_based_vuldetec_were_afraid_to_ask}. 

\subsection{Reporting omissions observed in practice}
Alongside author-stated limitations, we frequently encountered missing artifacts (code, prompts, preprocessing scripts) or incomplete reporting (splits, deduplication, compute, and inference settings), which obstructs verification, replication, and fair comparison. This is particularly salient for pipelines with non-trivial preprocessing (e.g., graph construction) and for hosted-LLM usage where model versions, context limits, and decoding choices may be under-specified \cite{research_239_2025_llm_based_vuldetec_were_afraid_to_ask,research_110_llm4cvd_short_results_directions}. To support reproducibility, we also maintain a companion GitHub index that aggregates, for each reviewed study, the publicly available artifacts explicitly linked by the authors (paper URLs and, when provided, code/repository links).

\paragraph{\textbf{Minimum reporting checklist for label-efficient CVD}}
We distill these recurring reporting omissions into a minimal checklist to improve comparability and reproducibility in future label-efficient CVD work.

\begin{itemize}
    \item \textbf{Task definition:} granularity (function/statement/line/file/project), label semantics (CWE\\/CVE/binary), and target domain.
    \item \textbf{Data protocol:} dataset versions, preprocessing (incl.\ deduplication), and split policy (project-wise, time-wise, cross-project); leakage controls.
    \item \textbf{Label budget:} unit (K-shot, fraction, counts), selection strategy, and any weak/implicit supervision used.
    \item \textbf{Representation + tooling:} tokenization, graph construction (if any), external analyzers, and failure cases of the toolchain.
    \item \textbf{Training/adaptation:} objectives, hyperparameters, early stopping, and robustness checks (seeds, ablations).
    \item \textbf{Inference/serving:} latency/throughput considerations where applicable; for hosted LLMs, model/version/date, decoding settings, context limits,
          and whether prompts/retrieval corpus are released.
    \item \textbf{Evaluation:} effort or imbalance-aware metrics where appropriate; cross-domain/cross-project validation if claiming generalization.
    \item \textbf{Artifacts:} code, scripts, prompts, and configs required to reproduce results.
\end{itemize}

\begin{rqsummary}{RQ5 — Summary}
\begin{enumerate}
    \item Author-stated limitations cluster around narrow benchmark coverage, limited external validity, operational constraints (compute/latency/hosted-LLM usage), and mechanism-linked brittleness.
    \item In addition, we observe recurring reporting omissions (artifacts, protocols, splits/deduplication, and inference details) that hinder verification and fair comparison; we distill these omissions into a minimal reporting checklist.
\end{enumerate}
\end{rqsummary}

\section{Threats to Validity}
\label{sec:threats_to_validity}

Our systematic mapping inherits standard validity threats in evidence synthesis: corpus coverage,
categorization/coding decisions, and limitations in primary-study reporting. Below we summarize the
main risks and the steps taken to reduce their impact.

\subsection{Search and Coverage Bias}
Our retrieval is CVD-centered but mechanism-driven: we iteratively expanded digital-library queries
from vulnerability terms to paradigm and representation keywords (e.g., contrastive/SSL/PU/few-shot,
prompting/in-context; AST/CFG/CPG/graph) and complemented this with backward/forward snowballing.
To keep the Main Set precise for CVD, we applied conservative first-pass screening criteria and
separated results into a \emph{Main} (CVD-primary) and an \emph{Inspiration} set (adjacent SE tasks with
methodologically relevant label-efficient mechanisms). Despite these mitigations, some relevant work
may be missed due to terminology variation, indexing limits, or screening decisions.

Our inclusion process may also capture \emph{grey literature} (e.g., technical reports or preprints not
yet peer-reviewed). We include such studies cautiously and use them primarily for mechanism
characterization and hypothesis generation, rather than as stand-alone support for strong comparative
claims. When included, we required clear methodological descriptions and situate these works relative
to peer-reviewed evidence (e.g., later peer-reviewed follow-ups, subsequent citations, or alignment in
experimental protocols), so that they contribute as structured inspiration while preserving a
conservative interpretation of the evidence base.

We restricted the corpus to English-language publications; evidence from non-English venues may be
underrepresented.

\subsection{Selection, Classification, and Coding Bias}
\textbf{Paradigm assignment.} Many pipelines blend mechanisms (e.g., contrastive pretraining with
supervised fine-tuning and consistency regularization~\cite{research_094_clever_multimodal_contrastive_cvd_representation, research_166_contrabert},
or active learning coupled with semi-supervised refinement~\cite{research_102_active_learining_CVD_pruning_bad_seeds}). In such cases, assigning a
single paradigm label requires judgment. We classified each study by its dominant contribution
(i.e., the main methodological driver emphasized by the paper) and recorded secondary roles in the
coding schema. This approach supports a readable synthesis, but may under-represent compositional
design details in individual studies.

\textbf{Main vs.\ Inspiration Sets.} Our Main/Inspiration distinction is a deliberate scoping choice:
Main studies treat code vulnerability detection as the primary task, whereas Inspiration studies
instantiate label-efficient paradigms on adjacent software engineering tasks (e.g., defect prediction,
clone detection) that are methodologically informative. This boundary improves clarity for CVD, but
may understate the extent to which results from adjacent domains transfer to vulnerability settings.

\textbf{Extraction uncertainty.} We extracted study attributes using a structured schema (supervision
regime, paradigm, representation, datasets/splits, metrics, label budgets, compute proxies, artifact
availability). However, budgets, negative sampling, and compute/serving cost are frequently reported
incompletely or only qualitatively. When details were missing, we used coarse categories (e.g.,
\emph{few-shot style} vs.\ \emph{percentage-labeled}) and marked fields as unknown rather than inferring
values. Interpretive fields (e.g., author-stated limitations and observed generalization/reporting
gaps) necessarily reflect reviewer judgment; we applied consistent criteria across the corpus and
revisited earlier entries once the taxonomy stabilized, but some subjectivity remains unavoidable.

\subsection{Reporting and Publication Bias}
Our conclusions are constrained by what studies report and what is likely to be published. In
particular, label budgets, compute cost, split construction, and artifact release are often omitted or
insufficiently specified, limiting verification and cross-paper comparison. Negative or inconclusive
findings may also be underrepresented. Consequently, our synthesis should be interpreted as
\emph{conservative}: where evidence is missing, we avoid strong claims, and we explicitly distinguish
between (i) limitations stated by authors and (ii) reporting omissions observed in the survey.

\subsection{Temporal Validity}
The ecosystem of code-oriented LLMs, tool-augmented pipelines, and evaluation benchmarks evolves
rapidly. Our search was last updated in \texttt{October 2025}; newer work may shift the balance of
paradigms, representations, and operational assumptions. To support updates, we provide the extraction
schema and analysis scripts in the replication package.


\section{Discussion and Research Agenda}
\label{sec:discussion}


Our findings across RQ1--RQ5 indicate that (i) label-efficient CVD is moving from isolated
techniques toward composable pipelines; (ii) structure-aware and domain-adaptive methods are
increasingly important for realistic deployments; and (iii) reporting practices around
label budgets, compute cost, and negative sampling remain inconsistent. We now outline a
prioritized research agenda grounded in these observations, followed by a constraint-first
decision guide for deployment.

\noindent To ground the discussion, Table~\ref{tab:design_map} consolidates our survey synthesis into a compact
\emph{design map}: paradigm families, their label-saving signals, typical CVD instantiations, representation
affinities, budget regimes, operational cost profiles, and recurring failure modes. We use this map to
derive a prioritized research agenda and a constraint-first practitioner guide.

\begin{table*}[t]
\centering
\scriptsize
\setlength{\tabcolsep}{4pt}
\renewcommand{\arraystretch}{1.15}

\caption{\textbf{Label-Efficient CVD Design Map (survey synthesis).}
Paradigm families, label-saving signals, typical instantiations in code vulnerability detection, representation affinity,
label-budget regimes, cost profiles, and common failure modes. Tok./Graph/Hybrid indicate representation affinity;
\textbf{Pre}/Train/\textbf{Inf} denote relative operational cost (L/M/H).}
\label{tab:design_map}

\newcolumntype{Y}{>{\raggedright\arraybackslash}X}

\begin{tabularx}{\textwidth}{X X X p{0.4cm} p{0.4cm} p{0.4cm} X p{0.4cm} p{0.4cm} p{0.4cm} X}
\toprule
\textbf{Paradigm family} &
\textbf{Label-saving signal} &
\textbf{Typical instantiation in CVD} &
\textbf{\raggedleft{Tok.}} &
\textbf{Gph.}&
\textbf{Hyb.} &
\textbf{Label-budget regime} &
\textbf{Pre} &
\textbf{Train} &
\textbf{Inf} &
\textbf{Common failure mode} \\
\midrule

\textbf{Semi-/Self-supervised} &
Pseudo-labels, consistency, self-training &
Teacher-student, uncertainty filtering, consistency regularization; sometimes graph-based propagation &
\cmark & \cmark & \cmark &
Often reported as a labeled fraction (small to moderate); definitions vary across studies &
\med & \high & \med &
Confirmation bias; noise amplification; sensitivity to split/leakage and negative sampling \\
\midrule
\textbf{Contrastive} &
View alignment (InfoNCE) &
Token augmentations; graph views (AST/CFG/DFG/CPG); domain-aligned contrastive pretraining &
\cmark & \cmark & \cmark &
Label-free pretraining + small supervised fine-tuning; label budget often implicit &
\high & \high & \med &
Brittle view/augmentation design; may wash out vulnerability cues; representation collapse \\
\midrule
\textbf{Few-shot / Meta-learning} &
Support set adaptation (MAML / Proto-style) &
Episodic training for new projects/CWEs; metric/prototypical variants; rapid adaptation settings &
\cmark & \xmark & \cmark &
Reported as $k$-shot (often 1-5 per CWE); not directly comparable as “\% of dataset” &
\low & \high & \med &
Highly sensitive to support-set quality and task construction; instability across episodes; poor calibration \\
\midrule
\textbf{Prompt tuning} &
Learn soft/hard/prefix prompts; freeze backbone &
Cloze reformulation; graph-aware prompt conditioning (prompts conditioning a GNN) &
\cmark & \xmark & \cmark &
Low-resource supervised (small labeled set) &
\low & \med & \med &
Template/verbalizer sensitivity; prompt overfitting; weak robustness under domain shift \\
\midrule
\textbf{LLM-based} &
Inference-time prompting/RAG/CoT or LoRA/adapters &
Few-shot prompts; RAG over CWE/CVE knowledge; adapter-based fine-tuning &
\cmark & \xmark & \cmark &
“Labels” become \textit{in-context} exemplars/demonstrations; rarely budgeted precisely &
\med & \med & \high &
Prompt/context sensitivity; truncation effects; non-determinism; evaluation leakage via retrieval/corpora \\

\bottomrule
\end{tabularx}

\end{table*}

\FloatBarrier

\subsection{From Single-Paradigm Methods to Composable Pipelines}
Recent work increasingly combines paradigms instead of treating them in isolation.
Multi-modal contrastive architectures couple structural and textual views~\cite{research_094_clever_multimodal_contrastive_cvd_representation};
semi-supervised pipelines leverage unlabeled code through teacher-student loops and
consistency regularization~\cite{research_74_2025_less_is_more_semi_supervised_deep_learning_vuldetec};
few-shot frameworks and prototype-based methods build on pretrained encoders for rapid
adaptation~\cite{research_76_2024_Few_VulD_few_shot_learning_framework_for_software_vuldetect, research_78_2020_VulMirror_few_shot_method_for_discovering_vulnerable_code_clone};
and prompt-based systems enrich LLMs with static-analysis signals and retrieval~\cite{research_74_2025_Structure_Enhanced_Prompt_Learning_Graph_Based_CodeVuldetect, research_162_expert_in_the_loop_cross_domain_in_domain_fsl_svd}.

The next step is to treat label-efficiency as a \emph{system property}, not a single technique:
end-to-end pipelines should make supervision, compute, and deployment constraints explicit, and then compose whatever mechanisms
are needed to meet them, rather than optimizing accuracy in a fixed, one-paradigm setup.

\subsection{Pillar A: Structure-aware and Multi-view Architectures}
\noindent\textit{Motivation from RQ1/RQ2:} token-only pipelines often underuse program structure, while
graph/hybrid views appear disproportionately in settings that demand robustness or fine-grained reasoning.

\paragraph{A1) Structure-aware LLMs rather than LLM-only.}
Token-only prompting may confuse patched and vulnerable variants or miss cross-function flows.
Injecting program structure, via graph-derived features, path summaries, or slices, can help
capture control and data dependencies. Structure-enhanced prompting and expert-in-the-loop LLM
systems suggest that combining token-level reasoning with structural context improves robustness~\cite{research_74_2025_Structure_Enhanced_Prompt_Learning_Graph_Based_CodeVuldetect, research_162_expert_in_the_loop_cross_domain_in_domain_fsl_svd}.

\paragraph{A2) Unified multi-view pretraining.}
Contrastive and multi-modal methods that align multiple representations
(e.g., tokens, graphs, and metadata) show promise~\cite{research_094_clever_multimodal_contrastive_cvd_representation, research_166_contrabert}.
A unified pretraining regime that jointly models code tokens, control/data-flow graphs,
and (when available) textual or binary artifacts could improve cross-domain generalization
and reduce labels needed for downstream adaptation.

\paragraph{A3) Budget-aware learning loops (Active $\times$ Semi/Self / PU).}
Supervision should be treated as a scarce resource. Active selection (uncertainty/diversity),
combined with teacher-student loops and consistency constraints, can focus human annotation on
hard cases~\cite{research_102_active_learining_CVD_pruning_bad_seeds, research_74_2025_less_is_more_semi_supervised_deep_learning_vuldetec}.
Standard practice should include label-efficiency curves and explicit accounting of (i) labeled
fractions/$K$-shot support, (ii) in-context exemplars, and (iii) unlabeled pool sizes. Where
reliable negatives cannot be guaranteed, PU-aware variants should be stated and stress-tested.

\subsection{Pillar B: Generalization Under Shift, Evolution, and Privacy}
\noindent\textit{Motivation from RQ3/RQ5:} transfer settings and distribution shift are common, but evidence is
often fragmented across datasets, projects, and CWE mixtures.

\paragraph{B1) Continual and few-shot CWE evolution.}
Few-shot frameworks demonstrate that some vulnerability types can be learned from very small
support sets~\cite{research_76_2024_Few_VulD_few_shot_learning_framework_for_software_vuldetect, research_78_2020_VulMirror_few_shot_method_for_discovering_vulnerable_code_clone, research_79_2025_Question_Answer_Methodology_for_vulnerable_source_code_review_via_prototype_based_MAML}.
However, long-tail CWEs and evolving patterns remain challenging. Continual learning with
parameter-efficient tuning, rehearsal, and drift-aware regularization is a natural extension.

\paragraph{B2) Robust cross-domain transfer with privacy.}
Cross-project and cross-domain studies reveal substantial distribution shifts between source
and target projects~\cite{research_100_CVD_cross_domain_graph_embedding_domain_adaptation, research_101_CVD_deep_domain_adaptation_max_margin_principle_cross_project, research_104_CSVD_TF}.
Future work should pursue domain/test-time adaptation that emphasizes causal invariants, as well as
federated or privacy-preserving regimes that respect code confidentiality while enabling shared
models~\cite{research_275_2025_transfer_learning_svd_using_transformer_models}.

\subsection{Pillar C: Closing the Loop From Detection to Repair}
\noindent\textit{Motivation from RQ1/RQ5:} most surveyed systems stop at detection, leaving patching and
verification underexplored in label-efficient settings.

\paragraph{C1) Tool-integrated detection and repair loops.}
Pipelines that connect detection to patch generation and verification are still rare~\cite{research_085_2025_DCodeBERT}.
Combining LLM-based patching with static analysis, concolic execution, and fuzzing can yield
end-to-end workflows that not only flag vulnerabilities but also propose and validate fixes
before integration.

\subsection{Pillar D: Evaluation Realism and Operational Cost}
\noindent\textit{Motivation from RQ4/RQ5:} evaluation and reporting choices (splits, negatives, cost) frequently
prevent fair comparison and overestimate deployability.

\paragraph{D1) Evaluation that matches reality.}
Evaluation practices often rely on random splits and accuracy/F1 metrics.
Future studies should standardize leakage-aware project-wise or temporal splits,
employ effort-aware metrics (e.g., E@20\%, Popt), and report calibration/abstention when
selective prediction is used. Negative sampling and class imbalance assumptions should be
documented explicitly (and sensitivity analyses reported when negatives may be noisy).

\paragraph{D2) Compute, cost, and carbon as first-class metrics.}
Given the diversity of pipelines; from graph-heavy detectors and autoencoders to LLM-based prompting;
compute and energy costs should be reported alongside accuracy~\cite{research_088_2024_Anvil, research_173_vulmae, research_252_2024_vuldetectbench, research_263_2024_enhancing_cvd_llm_prompt_engineering}.
Comparisons on ``accuracy per unit cost'' or ``bugs found per GPU-hour'' would help practitioners
choose methods that are both effective and practical.

\paragraph{Takeaway.}
Label-efficient CVD is evolving toward composable, deployment-oriented pipelines that integrate
structure-aware LLMs, multi-view representations, budget-aware learning loops, adaptation under
shift, and (increasingly) repair/verification. The most impactful progress is likely to come from
systems that jointly optimize supervision efficiency, operational cost, robustness, and evaluation realism.

\subsection{Practitioner decision guide}
\label{sec:decision_guide}

While RQ1--RQ5 characterize paradigms, budgets, costs, and threats to validity, practitioners often
face a constraint-first choice (label availability, negative reliability, domain shift, latency).
Table~\ref{tab:decision_guide} maps common deployment constraints to strategy families and their
dominant failure modes, based on patterns observed across the surveyed literature.

\begin{table}[t]
\centering
\small
\setlength{\tabcolsep}{4pt}
\renewcommand{\arraystretch}{1.15}
\caption{Practical decision guide for choosing label-efficient strategies in CVD under common constraints.}
\label{tab:decision_guide}
\begin{tabular}{p{3.1cm} p{3.4cm} p{5.6cm}}
\hline
\textbf{Constraint / scenario} & \textbf{Recommended family} & \textbf{Why + key caveat} \\
\hline
Very few labels (e.g., tens of labeled functions), need quick start
& LLM-based (in-context) \newline or Prompt tuning
& Fast deployment; uses demonstrations or light tuning. \textbf{Caveat:} prompt/context sensitivity; potential leakage via retrieval/corpora; report exemplars and context length. \\
\hline
Moderate labels (e.g., 5--20\%), stable pipeline desired
& Semi-/Self-supervised
& Exploits abundant unlabeled code when it matches the target distribution. \textbf{Caveat:} confirmation bias in pseudo-labeling; requires careful splits and filtering. \\
\hline
Strong domain shift (new project/org), want robustness
& Contrastive pretraining \newline + light fine-tuning
& Learns invariances from unlabeled data; can improve transfer. \textbf{Caveat:} view/augmentation design can erase vulnerability cues; report view construction. \\
\hline
Need adaptation across CWEs/tasks with few examples each
& Few-shot / Meta-learning
& Designed for rapid adaptation with small support sets. \textbf{Caveat:} highly sensitive to support quality and episode design; report $k$ and task sampling. \\
\hline
Limited compute at training time, but can afford light tuning
& Prompt tuning
& Parameter-efficient; cheaper than full fine-tuning. \textbf{Caveat:} template/verbalizer dependence; needs robustness checks across prompts. \\
\hline
\textbf{No reliable negatives / positives-only setting}
& \textbf{PU-aware training} \newline (often with semi-/self-supervision)
& Avoids assuming ``clean'' negatives; treats the remainder as unlabeled. \textbf{Caveat:} sensitive to prevalence and labeling bias; report assumptions and run sensitivity analyses. \\
\hline
\end{tabular}
\end{table}

\noindent\textbf{Operational guidelines.}
\begin{itemize}
  \item \textit{Negative reliability.} When reliable negatives cannot be guaranteed, methods that assume ``clean'' negative sampling should be avoided; PU-aware training or explicit unlabeled-positive assumptions are preferable, with sensitivity analyses.
  \item \textit{Latency/operational cost.} When CI or deployment latency is binding, LLM inference cost and graph extraction overhead should be treated as first-class metrics and reported as throughput/latency alongside predictive performance.
  \item \textit{Generalization under shift.} When cross-project generalization is the objective, evaluation should prioritize project-wise or temporal splits; random splits are likely to overestimate real-world gains.
\end{itemize}

\section{Conclusion}
\label{sec:conclusion}

This systematic mapping study provides a structured synthesis of label-efficient
strategies for CVD, and consolidates practical guidance on budget
reporting and evaluation. The literature has evolved from early deep sequence and program-graph
models to a diverse set of semi-/self-supervised, few-shot/meta-learning, contrastive, and
LLM-based approaches. By organizing these methods along a supervision spectrum, we characterize the
recurring mechanisms that leverage unlabeled code, transfer structural priors across projects,
generate pseudo-labels, learn from small support sets, and adapt to new domains or CWE types with
limited annotation.

Across RQ1--RQ5, four takeaways stand out. (1) Label-efficiency is increasingly achieved through
\emph{compositions} of mechanisms rather than isolated paradigms, combining pretraining, weak or
pseudo supervision, transfer, parameter-efficient adaptation, and retrieval-augmented prompting.
(2) Representation choices remain tightly coupled to what works in practice: token-based pipelines
naturally align with large-scale pretraining and prompting, while structure-aware views (e.g., graph
or flow signals) are frequently used when long-range dependencies and control/data relations matter;
hybrid designs are a recurring choice when robustness and transfer are prioritized. (3) Evidence is
hard to compare because label budgets and compute/serving cost are inconsistently specified; only a
subset of studies report budget sensitivity, effort-aware evaluation, or reproducible compute
details. (4) External validity remains fragile: distribution shift, near-duplicate leakage, label
noise, and uneven CWE coverage continue to limit robustness beyond a small set of well-curated
benchmarks.

Building on these observations, Section~\ref{sec:discussion} outlines a prioritized research agenda
and a constraint-first decision guide for practitioners, with emphasis on realistic splits,
de-duplication, effort-aware metrics, and transparent reporting of budgets, compute requirements,
and artifacts.

Overall, label-efficient learning will not replace expert labels or static/dynamic analysis, but it
can reduce annotation burden and improve scalability when paired with rigorous evaluation and
transparent reporting. We hope this study provides a clear conceptual framework and consolidated
evidence base that support the design of more robust and practical vulnerability detection systems.

\bibliographystyle{ACM-Reference-Format}
\bibliography{my_bibliography}

\end{document}